\documentstyle[12pt]{article}
  \input FEYNMAN
  \newlength{\extralineskip}
\newcounter{equnum}[section]
\def\theequnum{{\rm\thesection}.\arabic{equnum}}
\newcommand{\beq}{$$ \refstepcounter{equnum}}
\newcommand{\eeq}{\eqno (\theequnum) $$}
\newcommand{\bd}{\begin{displaymath}}
\newcommand{\ed}{\end{displaymath}}

\font\twlmsy=msbm10 at 12pt
\font\sevenmsy=msbm8
\font\fivemsy=msbm6
\newfam\Bbbfam
\textfont\Bbbfam=\twlmsy
\scriptfont\Bbbfam=\sevenmsy
\scriptscriptfont\Bbbfam=\fivemsy
\def\Bbb{\fam\Bbbfam}

\def\e{\, {\rm e}}

\def\tr{{\rm tr}}
\def\ps2{{\bar{\psi}\psi}}
\makeatletter
\newdimen\normalarrayskip              
\newdimen\minarrayskip                 
\normalarrayskip\baselineskip
\minarrayskip\jot
\newif\ifold             \oldtrue            \def\new{\oldfalse}
\def\arraymode{\ifold\relax\else\displaystyle\fi} 
\def\@arrayskip{\ifold\baselineskip\z@\lineskip\z@
     \else
     \baselineskip\minarrayskip\lineskip2\minarrayskip\fi}
\def\@arrayclassz{\ifcase \@lastchclass \@acolampacol \or
\@ampacol \or \or \or \@addamp \or
   \@acolampacol \or \@firstampfalse \@acol \fi
\edef\@preamble{\@preamble
  \ifcase \@chnum
     \hfil$\relax\arraymode\@sharp$\hfil
     \or $\relax\arraymode\@sharp$\hfil
     \or \hfil$\relax\arraymode\@sharp$\fi}}
\def\@array[#1]#2{\setbox\@arstrutbox=\hbox{\vrule
     height\arraystretch \ht\strutbox
     depth\arraystretch \dp\strutbox
     width\z@}\@mkpream{#2}\edef\@preamble{\halign \noexpand\@halignto
\bgroup \tabskip\z@ \@arstrut \@preamble \tabskip\z@ \cr}%
\let\@startpbox\@@startpbox \let\@endpbox\@@endpbox
  \if #1t\vtop \else \if#1b\vbox \else \vcenter \fi\fi
  \bgroup \let\par\relax
  \let\@sharp##\let\protect\relax
  \@arrayskip\@preamble}
\makeatother

\begin{document}

\begin{titlepage}

\baselineskip=12pt

\rightline{OUTP-97-09P}
\rightline{hep-th/9703071}
\rightline{   }
\rightline{\today}

\vskip 0.6truein
\begin{center}

\baselineskip=24pt
{\Large\bf Liouville Dressed Weights and Renormalization of Spin in
Topologically Massive Gravity}\\
\baselineskip=12pt
\vskip 0.8truein
{\bf Ian I. Kogan} and {\bf Richard J. Szabo}\\

\vskip 0.3truein

{\it Department of Physics -- Theoretical Physics\\ University of Oxford\\ 1
Keble Road, Oxford OX1 3NP, U.K.}\\

\vskip 1.3 truein

\end{center}

\begin{abstract}

\baselineskip=12pt

We examine the relations between observables in two- and three-dimensional
quantum gravity by studying the coupling of topologically massive gravity to
matter fields in non-trivial representations of the three-dimensional Lorentz
group. We show that the gravitational renormalization of spin up to one-loop
order in these theories reproduces the leading orders of the KPZ scaling
relations for quantum Liouville theory. We demonstrate that the two-dimensional
scaling dimensions can be computed from tree-level Aharonov-Bohm scattering
amplitudes between the charged particles in the limit where the
three-dimensional theory possesses local conformal invariance. We show how the
three-dimensional description defines scale-dependent weights by computing the
one-loop order anomalous magnetic moment of fermions in a background
electromagnetic field due to the renormalization by topologically massive
gravity. We also discuss some aspects concerning the different phases of
three-dimensional quantum gravity and argue that the topological ones may be
related to the branched polymer phase of two-dimensional quantum gravity.

\end{abstract}

\end{titlepage}

\clearpage\newpage

\baselineskip=18pt

\section{Introduction}

It has been known for quite some time that gauge and gravity theories in a
three-dimensional spacetime can have dynamics which are not possible in other
dimensions. Topologically massive gravity \cite{djt} is the modification of
three-dimensional Einstein gravity by the addition of the gravitational
Chern-Simons term to the usual Einstein-Hilbert action. In contrast to the
ordinary Einstein theory, which is non-dynamical due to the equivalence of the
Einstein and Riemann tensors in three dimensions, it has a propagating, massive
graviton degree of freedom. It has been shown \cite{desyang,kesz} to be the
unique locally dynamical and unitary gravity model which is power-counting
renormalizable, and it exhibits a variety of novel effects such as a
gravitational analog of the Aharonov-Bohm effect \cite{desmc,a-cksgrav},
violation of the equivalence principle, and anti-gravity \cite{desmc}. In this
Paper we will discuss the relationship between topologically massive gravity
and the gravity sector of string theory (Liouville theory).

The relation between topologically massive gravity on a three-dimensional
spacetime manifold $\cal M$ with boundary $\partial{\cal M}$ and
two-dimensional gravity on $\partial{\cal M}$ was first conjectured in
\cite{kogan91}, and then derived by Carlip in \cite{carlip1} from a formal
path-integral approach. The induced Liouville gravity theory on $\partial\cal
M$ is \cite{carlip1,kogan1}
\beq
S_L[\phi]=\int_{\partial{\cal
M}}d^2z~\left(k'\partial_{\bar{z}}\phi\partial_{z}\phi
- QR^{(2)}\phi + \Lambda \e^{\alpha_+\phi} \right)
\label{liouville}\eeq
where $\phi$ is the induced dilaton field, the two-dimensional cosmological
constant $\Lambda$ is equal to the square of the three-dimensional topological
graviton mass \cite{kogan1}, and $k'$ is the gravitational Chern-Simons
coefficient which is related to the central charge $c$ of the $SL(2,{\Bbb R})$
Kac-Moody algebra by $c=-k'-4$. This connection expands the general
relationship between three-dimensional topological field theory and
two-dimensional conformal field theory \cite{3d2d}, and it describes
gravitational dressing effects in the topological membrane formulation of
string theory in which the string world-sheet is filled in and viewed as the
boundary of a 3-manifold \cite{kogan2,carkog1}.

In the general relationship between topological gauge theories and conformal
field theories, one can describe the fundamental quantum observables of the
primary conformal fields by coupling the gauge theory to charged matter. The
$n$-point correlation functions of the conformal field theory can be decomposed
\beq
\left\langle\prod_{i=1}^nV(z_i,\bar
z_i)\right\rangle=\left\langle\prod_{i=1}^nV_{\rm
L}(z_i)\right\rangle\left\langle\prod_{i=1}^nV_{\rm R}(\bar z_i)\right\rangle
\label{CFTcorr}\eeq
in terms of products of left and right conformal blocks, where $V_{\rm L}(z_i)$
and $V_{\rm R}(\bar z_i)$ are the holomorphic and anti-holomorphic chiral
vertex operators corresponding to the left-right symmetric vertex operator
$V(z_i,\bar z_i)$. In the corresponding three-dimensional gauge theory we
consider the 3-manifold ${\cal M}=\Sigma\times[0,1]$ whose two boundaries
$\Sigma_{\rm L}$ and $\Sigma_{\rm R}$ are connected by a finite interval. A
Chern-Simons gauge theory in $\cal M$ induces both left- and right-moving
sectors of the two-dimensional conformal field theory, and an insertion of a
vertex operator on the Riemann surface $\Sigma$ is equivalent to insertions of
the chiral vertex operators $V_{\rm L}(z)$ and $V_{\rm R}(\bar z)$ on the left-
and right-moving worldsheets $\Sigma_{\rm L}$ and $\Sigma_{\rm R}$,
respectively. The insertions corresponding to the correlation functions
(\ref{CFTcorr}) are induced by path-ordered products of open Wilson line
operators \cite{3d2d}--\cite{a-cks}
\beq
W_{{\cal C}_{z_1,\bar z_1};\dots;{\cal C}_{z_n,\bar
z_n}}^{(R_1,\dots,R_n)}[A^{(1)},\dots,A^{(n)}]=\prod_{i=1}^n~\mbox{tr}_{R_i}
{}~P\exp\left(i\int_{{\cal C}_{z_i,\bar z_i}}A^{(i)}\right)
\label{wilsonvert}\eeq
along the oriented paths ${\cal C}_{z_i,\bar z_i}\subset{\cal M}$ with
endpoints $z_i\in\Sigma_{\rm L}$ and $\bar z_i\in\Sigma_{\rm R}$. Correlators
of insertions of the Wilson lines (\ref{wilsonvert}) in $\cal M$ induce phase
factors from adiabatical rotation of charged particles coupled to the gauge
fields $A^{(i)}$ in the representations $R_i$ of a gauge group $G$ (this set of
quantum numbers depends on the types of vertex operators under consideration).
The quantum particles propagate along ${\cal C}_{z_i,\bar z_i}$ from left- to
right-moving worldsheets, so that the corresponding linking of the Wilson lines
from the adiabatical rotations in $\cal M$ are equivalent to braidings of the
associated vertex operators on $\Sigma$ whose induced phases correspond to the
conformal dimensions of the primary fields.

In the connection between three-dimensional gravity and Liouville theory, this
approach can therefore provide geometrical and dynamical descriptions of the
characteristics of two-dimensional quantum gravity within the more conventional
framework of quantum field theory. This feature has been exploited recently in
\cite{a-cks} where it was demonstrated that the anomalous scaling dimensions of
primary operators in the induced conformal field theories can be calculated
perturbatively in the three-dimensional quantum field theory by viewing the
induced spin as the Aharonov-Bohm phase factor \cite{ab} that arises in the
scattering amplitudes between dynamical charged particles interacting via
exchange of the topologically massive bosons. In topologically massive gravity,
the Liouville-dressing of vertex operators on the worldsheet can be expected to
coincide with graviton exchanges between charged particles in the bulk. It was
argued in \cite{a-cks} that this identification reproduces the leading orders
of the Knizhnik-Polyakov-Zamolodchikov (KPZ) scaling relations \cite{kpz}
\beq
\Delta-\Delta_0=\frac{\Delta(1-\Delta)}{c+2}
\label{KPZtransf}\eeq
for quantum Liouville theory. The formula (\ref{KPZtransf}) describes the
transformation of the bare spin $\Delta_0$ of some primary conformal fields to
the spin $\Delta$ due to the dressing by two-dimensional quantum gravity. The
iterative large-$k'$ expansion of (\ref{KPZtransf}) for the branch which has
$\Delta(\Delta_0=0)=0$ is
\beq
\Delta_-=\Delta_0+\frac{\Delta_0-\Delta_0^2}{k'}+
\frac{2\Delta_0^3-\Delta_0-\Delta_0^2}{k'^2}+\dots
\label{KPZexp}\eeq
It was shown in \cite{a-cks} that when three-dimensional charged scalar matter
fields, with an anomalous fractional spin $\Delta_0$ from their coupling to a
Chern-Simons gauge field, are coupled to topologically massive gravity, the
one-loop order scattering amplitudes reproduce the leading orders of
(\ref{KPZexp}). This establishes a non-trivial direct link between the
observables of the two- and three-dimensional gravity theories. The
three-dimensional description of non-critical string theory in this way
suggests a clearer, geometrical picture of its observables from a dynamical
point of view, and it also provides evidence that a suitably defined quantum
measure does indeed exist for the path-integral of the Liouville gravity
theory.

In this Paper we shall explore further the relationship between
three-dimensional gravitational scattering amplitudes and the gravitational
dressing of conformal weights as predicted by the KPZ formula (\ref{KPZexp}),
emphasizing the roles played by the various symmetry groups of the
three-dimensional theory. We will study the coupling of topologically massive
gravity to matter fields which carry a non-trivial representation of the local
Lorentz group of the three-dimensional spacetime, wherein one expects to be
able to analyse tree-level amplitudes because of the non-zero bare spin that
the fields possess. We shall see that the tree-level renormalization of spin is
controlled by the interactions with the spin-connection field of topologically
massive gravity. This analysis illustrates how the various fields of
topologically massive gravity conspire in general to dress the bare observables
in such a way so as to reproduce the KPZ formula (\ref{KPZtransf}). In this way
we obtain a dynamical picture for the geometric realization of the mysterious
``hidden" $SL(2,{\Bbb R})\cong SO(2,1)$ Kac-Moody symmetry group of Liouville
theory \cite{kpz} as the local Lorentz group of three-dimensional gravity. As
we shall see, the interpretation of these dressed spins as Aharonov-Bohm phases
requires tuning the topological graviton mass to that point where the theory
has local three-dimensional conformal invariance and defines a topological
$SO(2,1)$ Chern-Simons gauge theory.

In light of this requirement, we consider also the anomalous spin through the
gravitational renormalization of the magnetic moment of fermion fields coupled
to a background electromagnetic field. We show that, at one-loop order, this
yields an unambiguous definition of the Liouville dressed spins which also
agrees with the two-dimensional theory. It does, however, require taking the
limit where the topologically massive gravity model reduces to ordinary
Einstein gravity (the same limit that was required in \cite{a-cks}), i.e.
tuning the topological graviton mass to the point where the theory becomes a
topological $ISO(2,1)$ Chern-Simons gauge theory. We will see that the
determination of the weights at this $ISO(2,1)$-invariant point is equivalent
to an identification of this limit of topologically massive gravity with
three-dimensional topological gravity. This then defines the dressed spins
(\ref{KPZexp}) in the three-dimensional theory for all scales. We shall see in
fact that the three-dimensional approach, in contrast to the Liouville theory,
defines a scale-dependent conformal dimension through such higher-loop order
renormalizations of the spin of the gravity-coupled fermion fields. These
results may also shed light on how the ghost sector of Liouville theory is
realized in terms of the $SO(2,1)$ spin-connection in topologically massive
gravity \cite{carkog2}.

The organization of this Paper is as follows. In Section 2 we review the
first-order formalism for topologically massive gravity and write down its
Feynman rules. In Section 3 we briefly review the perturbative approach to
determining conformal dimensions and argue heuristically how this will
reproduce the leading orders of the KPZ formula when the gravity theory is
coupled to spinning matter fields. In Section 4 we verify these arguments by
explicit tree-level calculations for a few cases and point out some intriguing
features of the amplitudes. In Section 5 we compute the one-loop order
renormalization of the fermionic magnetic moment and discuss the scale
dependence of the weights that are so derived. Finally, in Section 6 we discuss
the possibility of connecting different phases of topologically massive
gravity, including the one that admits a spacetime with a conical singularity
(and hence a gravitational Aharonov-Bohm effect), with models of
two-dimensional quantum gravity coupled to $c>1$ conformal matter fields. In
these cases the perturbative approach breaks down, so that the apparent mystery
of the two-dimensional theories in these instances are evident in the
three-dimensional gravity models as well. An Appendix at the end of the Paper
summarizes some of the identities used for the more complicated calculations.

\section{Topologically Massive Gravity in the First Order Formalism}

We begin by briefly reviewing the basic properties of topologically massive
gravity and its Feynman rules. We will be interested in describing this model
within the first order formalism for general relativity \cite{deserx}. For
this, we introduce the dreibein fields $e^a=e_\mu^adx^\mu$ which span the frame
bundle of the oriented three-dimensional spacetime manifold $\cal M$ which has
metric of Minkowski signature. Here and in the following greek letters will
label the spacetime indices (i.e. the components of the local basis vectors of
the tangent space) and latin letters will denote the basis indices of the local
Lorentz group $SO(2,1)$ of the tangent bundle. The dreibein fields are related
to the metric $g$ of $\cal M$ by the orthonormality condition
$g^{\mu\nu}e_\mu^ae_\nu^b=\eta^{ab}={\rm diag}(1,-1,-1)$, or equivalently by
the completeness relation
\beq
\eta_{ab}~e^a\otimes e^b=g
\label{dreicompl}\eeq

The action for topologically massive gravity in the first order formalism is
\beq
S_{TMG}=S_E+S_{CS}^{({\rm grav})}+S_\lambda
\label{TMGaction}\eeq
where
\beq
S_{CS}^{({\rm grav})}=\frac{k'}{8\pi}\int_{\cal M}\left(\omega^a\wedge
d\omega^a+\frac{2}{3}\epsilon^{abc}\omega^a\wedge\omega^b\wedge\omega^c\right)
\label{csgrav}\eeq
is the parity-odd gravitational Chern-Simons action, and
$\omega^a=\epsilon^{abc}\omega^{bc}$ with
$\omega^{ab}=-\omega^{ba}=\omega_\mu^{ab}dx^\mu$ the spin-connection of the
frame bundle of $\cal M$. We use the convention $\epsilon^{012}=+1$ for the
antisymmetric tensor, and we shall always assume that the gravitational
Chern-Simons coefficient $k'\in{\Bbb R}$ is large enough so that perturbation
theory makes sense. The Einstein-Hilbert action is
\beq
S_E=\kappa\int_{\cal M}e^a\wedge R^a
\label{einsteinaction}\eeq
where
\beq
R^a=R^a_{\mu\nu}dx^\mu\wedge dx^\nu=d\omega^a+\epsilon^{abc}\omega^b\wedge
\omega^c
\label{curvom}\eeq
is the curvature of the spin-connection $\omega^a$, and $\kappa$ is the Planck
mass. We have also included in (\ref{TMGaction}) the term
\beq
S_\lambda=\int_{\cal M}\lambda^a\wedge\left(de^a+2\epsilon^{abc}\omega^b\wedge
e^c\right)
\label{lagrangeterm}\eeq
where $\lambda^a=\lambda_\mu^adx^\mu$ are Lagrange multiplier fields that
enforce the constraint which ensures that the usual covariant derivative
$\nabla$ constructed from the spin-connection is compatible with the metric $g$
(i.e. $\nabla e^a=0$) so that the Einstein-Hilbert action in the first order
formalism coincides with the usual one of general relativity. This means that
we are working in the minimal formalism for general relativity in which
$\omega$ is the Levi-Civita connection of the spin bundle of $\cal M$.

The ordinary, pure Einstein theory (\ref{einsteinaction}) in three-dimensions
has no propagating degrees of freedom and is a topological field theory. In
fact, it can be regarded as a topological Chern-Simons gauge theory with gauge
group defined by the Poincar\'e group $ISO(2,1)\supset SO(2,1)$ of the
spacetime \cite{Town}. The addition of the gravitational Chern-Simons term
(\ref{csgrav}) makes the gravitons of the theory massive with topological mass
\beq
M=8\pi\kappa/k'
\label{gravmass}\eeq
In the infrared limit $M\to\infty$ (equivalently $k'\to0$) this propagating
degree of freedom decouples and the Chern-Simons term in (\ref{TMGaction})
becomes irrelevant. Then the kinetic term in the induced Liouville action
(\ref{liouville}) vanishes and the gravity theory (\ref{TMGaction}) induces the
topological field theory on $\partial{\cal M}$ which describes the moduli space
of Riemann surfaces. This was first argued in \cite{Town} where it was observed
that the physical phase space for three-dimensional Einstein gravity is closely
related to the moduli space of complex structures. It was subsequently shown in
\cite{carkog1} that the only contribution to the gauge-fixed path integral over
three-dimensional metrics in the interior of $\cal M$ is from an integration
over the moduli space of $\partial{\cal M}$ with the same integration measure
(the Weil-Petersson measure) that naturally arises in string theory. The
resulting quantum mechanics on moduli space was also studied in \cite{carkog1}.
For finite $M$, the world-sheet and bulk scales are related by
\beq
\Lambda=M^2
\label{LM2}\eeq
which establishes a non-trivial correspondence between the quantum Liouville
theory and topologically massive gravity. Some values of the world-sheet
cosmological constant $\Lambda$ then correspond to different symmetries of the
three-dimensional theory. Various physical applications of this feature, for
example to $SL(2,{\Bbb R})/U(1)$ black holes and $c=1$ strings, are discussed
in \cite{kogan1}, and the quantum mechanics on moduli spaces in the full
topologically massive gravity theory is studied in \cite{kogan91}.

The inclusion of the gravitational Chern-Simons term regulates the ultraviolet
divergences of the pure Einstein gravitational field theory. The effective
coupling constant of the topologically massive gravity theory is the
super-renormalizable, dimensionless expansion parameter $M/\kappa\sim1/k'$. The
topological graviton mass also regulates infrared divergences of the pure
Einstein theory. The action (\ref{TMGaction}) is diffeomorphism invariant (i.e.
generally covariant) and it also possesses a local $SO(2,1)$-invariance defined
by rotations of the dreibein fields. The spin-connection is a gauge field of
the local Lorentz group, and (\ref{csgrav}) can be regarded as a Chern-Simons
action for an $SO(2,1)$ gauge theory with connection $\omega$. Both the
dreibein and Lagrange multiplier fields above transform in the adjoint
representation of this gauge group. With this point of view, we will study the
model (\ref{TMGaction}) perturbatively by expanding the graviton field about a
flat background metric. This can be done in the ``perturbative phase" of the
theory wherein $\langle e_\mu^a\rangle=\delta_\mu^a$ \cite{kogan1}. Then we
shift the dreibein fields as
\beq
e_\mu^a\to e_\mu^a+\delta_\mu^a
\label{dreishift}\eeq
and view the topologically massive gravity theory as a quantum field theory on
a flat space. Introducing the new variables $\beta^a$ defined by
\beq
\beta^a=\lambda^a+\kappa\omega^a
\label{betadef}\eeq
the topologically massive gravity action becomes
\beq\new{\begin{array}{c}
S_{TMG}=\int d^3x~\left\{\epsilon^{\mu\nu\lambda}\beta_\mu^a\partial_\nu
e_\lambda^a+2(\beta_\mu^\mu\omega_\nu^\nu-\beta_\nu^\mu\omega_\mu^\nu)+
\frac{k'}{8\pi}\epsilon^{\mu\nu\lambda}\omega_\mu^a\partial_\nu\omega_
\lambda^a-\kappa(\omega^\mu_\mu\omega_\nu^\nu-\omega_\nu^\mu\omega_\mu^\nu)
\right.\\\left.+\epsilon^{\mu\nu\lambda}\epsilon^{abc}\left(2\beta_\mu^a
\omega_\nu^be_\lambda^c-\kappa e_\mu^a\omega_\nu^b\omega_
\lambda^c+\frac{k'}{12\pi}\omega_\mu^a\omega_\nu^b\omega
_\lambda^c\right)\right\}\end{array}}
\label{TMGshift}\eeq
where here and in the following all repeated indices are understood to be
summed over by contracting with the flat Minkowski metric $\eta^{ab}$. We shall
ignore the gauge-fixing and ghost field terms which are irrelevant for what
follows.

The Feynman rules for topologically massive gravity in the first order
formalism can now be read off from the action (\ref{TMGshift}). In the
following we shall work in the transverse Landau gauge. The $e\beta$ and
spin-connection propagators are \cite{a-cks}
\beq
\left\langle\!\!\left\langle\beta_\mu^i(p)e_\nu^j(-p)\right\rangle\!\!\right
\rangle=\eta^{ij}~
\frac{\epsilon_{\mu\nu\lambda}p^\lambda}{p^2}~~~~~,~~~~~
\left\langle\!\!\left\langle\omega_\mu^i(p)\omega^j_\nu(-p)\right\rangle
\!\!\right\rangle=\frac{4\pi i}{k'}~\Omega_{\mu\nu}^{ij}(p)
\label{omprop}\eeq
where the averages $\langle\!\langle\cdot\rangle\!\rangle$ denote free field
Gaussian correlators in momentum space, and
\beq\new{\begin{array}{lll}
\Omega_{\mu\nu}^{ij}(p)&=&\frac{M(\Lambda_\mu^i(p)\Lambda_\nu^j(p)-\Lambda
_{\mu\nu}(p)\Lambda^{ij}(p)-\Lambda_\mu^j(p)\Lambda_\nu^i(p))}{2(p^2-M^2)}
-\frac{(\Lambda_{\mu\nu}(p)\Lambda^{ij}(p)-\Lambda_\mu^j(p)\Lambda
_\nu^i(p))}{2M}\\&
&-\frac{1}{2M^3}(\Lambda_{\mu\nu}(p)p^ip^j+\Lambda^{ij}(p)p_\mu
p_\nu-\Lambda_\mu^j(p)p^ip_\nu-\Lambda_\nu^i(p)p_\mu p^j)
\\& &-\frac{i(\epsilon_{\mu\nu\lambda}p^\lambda\Lambda^{ij}(p)+
\epsilon^{ijk}p_k\Lambda_{\mu\nu}(p))}{2(p^2-M^2)}\\& &-\frac{i}{2M^2}
\left(\epsilon^{i\lambda}_{~~\mu}p_\lambda\frac{p_\nu
p^j}{p^2}+\epsilon_\nu^{~\lambda j}p_\lambda\frac{p_\mu
p^i}{p^2}+\epsilon_{\mu\lambda\nu}p^\lambda\frac{p^i
p^j}{p^2}-\epsilon^{i\lambda j}p_\lambda\frac{p_\mu
p_\nu}{p^2}\right)\end{array}}
\label{Xisoln}\eeq
with
\beq
\Lambda_{\mu\nu}(p)=\eta_{\mu\nu}-p_\mu p_\nu/M^2
\label{Pidef}\eeq

The pure graviton and $\omega e$ propagators can then be obtained from the
momentum space convolutions
\beq\new{\begin{array}{lll}
D_{\mu\nu}^{ij}(p)&\equiv&\langle\!\langle
e_\mu^i(p)e_\nu^j(-p)\rangle\!\rangle\\&=&-4\langle\!\langle
e_\mu^i(p)\beta_\lambda^k(-p)\rangle\!\rangle~[\Sigma^{kl}]_{\lambda\rho}~
\langle\!\langle\omega_\rho^l(p)\omega_\alpha^m(-p)\rangle\!\rangle
{}~[\Sigma^{mn}]_{\alpha\sigma}~\langle\!\langle\beta_\sigma^n(p)e_\nu^j(-p)
\rangle\!\rangle\\E_{\mu\nu}^{ij}(p)&\equiv&\langle\!\langle\omega^i_
\mu(p)e_\nu^j(-p)\rangle\!\rangle~=~2i\langle\!\langle\omega^i_\mu(p)\omega_
\lambda^k(-p)\rangle\!\rangle ~[\Sigma^{kl}]_{\lambda\rho}~\langle\!\langle
\beta_\rho^l(p)e_\nu^j(-p)\rangle\!\rangle
\end{array}}
\label{Ddef}\eeq
where
\beq
[\Sigma^{kl}]_{\lambda\rho}
=\delta_\lambda^k\delta^l_\rho-\delta_\rho^k\delta^l_\lambda
\label{Xdef}\eeq
is the spin matrix generator on the space of spin-1 fields. This yields
\cite{a-cks}
\beq\new{\begin{array}{lll}
D_{\mu\nu}^{ij}(p)&=&\frac{i}{\kappa}\left(\frac{M^2}{2p^2(p^2-M^2)}
\left\{\left(\frac{p^2}{M^2}-2\right)\eta_{\mu\nu}^\perp(p)\eta^{\perp
ij}(p)+\delta_\mu^{\perp i}(p)\delta_\nu^{\perp j}(p)+\delta_\mu^{\perp
j}(p)\delta_\nu^{\perp i}(p)\right\}\right.\\&
&~~~\left.+\frac{iM}{4}\frac{p^\lambda}{p^2(p^2-M^2)}\left\{\epsilon_{\mu~
\lambda}^{~i}\delta_\nu^{\perp
j}(p)+\epsilon_{\mu~\lambda}^{~j}\delta_\nu^{\perp
i}(p)+\epsilon_{\nu~\lambda}^{~i}\delta_\mu^{\perp
j}(p)+\epsilon_{\nu~\lambda}^{~j}\delta_\mu^{\perp
i}(p)\right\}\right)\\E_{\mu\nu}^{ij}(p)&=&-\frac{8\pi
i}{k'M}\left(\frac{i\epsilon_{\mu\nu\lambda}p^\lambda}{p^2}\eta^{ij}-\Omega_
{\mu\lambda}^{ij}(p)\delta^{\perp\lambda}_\nu(p)\right)\end{array}}
\label{gravprop}\eeq
where
\beq
\eta_{\mu\nu}^\perp(p)=\eta_{\mu\nu}-p_\mu p_\nu/p^2
\eeq
is the symmetric, transverse projection operator on the momentum space of
vectors. The Green's function $D_{\mu\nu}^{ij}(p)$ is the usual Deser-Yang
graviton propagator \cite{desyang}.

\section{Anomalous Spin in Topologically Massive Gravity}

We shall now outline the method, described in detail in \cite{a-cks}, for
identifying the conformal dimensions of primary operators in two-dimensional
conformal field theory as the transmuted spins that appear in charged particle
scattering amplitudes in three-dimensional perturbative gauge theory. The
minimal coupling of a Chern-Simons gauge field $A=A_\mu^aT^adx^\mu$ to a
conserved matter current $J^\mu=J^\mu_aR^a$ is described by the
parity-violating action
\beq\new{\begin{array}{lll}
S_{CS}^{[G]}+S_J&=&\int_{\cal M} {k\over4\pi}~\tr\left(A\wedge dA +
{2\over3}A\wedge A\wedge A\right)+\int_{\cal M}2~{\rm tr}~A\wedge\star
J\\&=&\int_{\cal
M}d^3x~\frac{k}{4\pi}\epsilon^{\mu\nu\lambda}~\tr\left(A_\mu\partial_\nu
A_\lambda+\frac{2}{3}A_\mu A_\nu A_\lambda\right)+\int_{\cal
M}d^3x~\sqrt{g}~2~{\rm tr}~J^\mu A_\mu^aR^a\end{array}}
\label{csaction}\eeq
where $R[G]$ is a unitary irreducible representation of the semi-simple gauge
group $G$ whose Hermitian generators $T^a$ are normalized as
$\tr~T^aT^b=-\frac12\delta^{ab}$. Again we omit the gauge-fixing and ghost
field terms which play no role in the following. The momentum space bare gluon
propagator from (\ref{csaction}) in the transverse Landau gauge is
\beq
{\cal G}^{ab}_{\mu\nu}(p)\equiv\left\langle\!\!\left\langle
A_\mu^a(p)A_\nu^b(-p)\right\rangle\!\!\right\rangle=-\frac{4\pi}{k}
\delta^{ab}\frac{\epsilon_{\mu\nu\lambda}p^\lambda}{p^2}
\label{puregaugeprop}\eeq

The conformal dimension $\Delta$ of a primary operator in the induced conformal
field theory on $\partial{\cal M}$ can then be determined as the transmuted
spin factor that appears in the invariant amplitude for the scattering of two
charged particles, of initial momenta $p_1$ and $p_2$ represented by the
current $J$, in the infrared (non-relativistic) regime $q^2\to0$ of the quantum
field theory (Fig. 1),
\beq
{\cal A}(p_1,p_2;q)\equiv i~\tr~J^\mu(2p_1-q){\cal
G}_{\mu\nu}(q)J^\nu(2p_2+q)=-16\pi
i\dim(G)\Delta(k)\frac{\epsilon_{\mu\nu\lambda}p_1^\mu p_2^\nu q^\lambda}{q^2}
\label{abrelgen}\eeq
where $q$ is the momentum transfer, and
\beq
\Delta(k)=T_R[G]\sum_{n\geq0}\frac{\Delta^{(n)}}{k^n}
\label{dimexpk}\eeq
is the anomalous spin of the charged particles induced by their interaction
with the Chern-Simons gauge field $A$. It can be computed order by order
perturbatively in the Chern-Simons coupling constant $1/k$. The coefficients
$\Delta^{(n)}$ of the expansion (\ref{dimexpk}) depend only on invariants of
the gauge group $G$ and of the local Lorentz group of the spacetime, and
\beq
T_R[G]~{\bf1}=\sum_{a=1}^{\dim G}R^aR^a
\label{Rcasimir}\eeq
is the quadratic Casimir operator of the Lie group $G$ in the representation
$R[G]$. In the (non-relativistic) center of momentum frame, the amplitude
(\ref{abrelgen}) coincides with the usual Aharonov-Bohm amplitude for the
scattering of a charge off of a flux \cite{a-cks,ab,dekss} which is observable
as a long-ranged effect in the theory.

\begin{figure}
\begin{picture}(30000,10000)
\small
\put(22000,3500){\makebox(0,0){$q$}}
\put(15000,1000){\makebox(0,0){$p_2$}}
\put(28000,1000){\makebox(0,0){$p_2+q$}}
\put(15000,6000){\makebox(0,0){$p_1$}}
\put(28000,6000){\makebox(0,0){$p_1-q$}}
\drawline\fermion[\E\REG](21000,1000)[5000]
\drawline\fermion[\W\REG](21000,1000)[5000]
\drawline\photon[\N\REG](21000,1000)[5]
\drawline\fermion[\E\REG](\photonbackx,\photonbacky)[5000]
\drawline\fermion[\W\REG](\photonbackx,\photonbacky)[5000]
\put(21000,1000){\circle*{500}}
\put(21000,6000){\circle*{500}}
\end{picture}
\begin{description}
\small
\baselineskip=12pt
\item[Figure 1:] The scattering amplitude for two charged particles. Here
$p_1,p_2$ denote the incoming particle momenta and $q$ is the momentum
transfer. Straight lines denote the external charged matter fields, wavy lines
depict the gauge fields, and the solid circles represent the minimal coupling
of the particle current $J^\mu$ to the gluon field.
\end{description}
\end{figure}

It was demonstrated in \cite{a-cks} that the leading order, tree-level
amplitude in the case where $J$ represents charged scalar matter fields yields
\beq
\Delta^{(1)}=1
\label{Delta1CS}\eeq
{}From this result we can naively suggest how the effect of the spin connection
in the topologically massive gravity action can reproduce the leading orders of
the KPZ formula (\ref{KPZexp}). We are interested in coupling the gravity
theory of the previous Section to some spinning matter fields. For this, we
consider a dynamical $(2j+1)$-component field $\Phi^{(j)}$ in the unitary
irreducible spin-$j$ representation of the local Lorentz group $SO(2,1)$ of the
spacetime, where $2j\in{\Bbb Z}^+$. According to (\ref{KPZexp}) the bare spin
$\Delta_0=j$ of these fields should renormalize as
\beq
\Delta_j(k')=j-j(j-1)/k'+j(2j^2-1-j)/k'^2+\dots
\label{KPZj}\eeq
as a result of the gravitational dressing in the induced Liouville gravity
theory on $\partial{\cal M}$.

As mentioned before, the spin-connection $\omega_\mu^i$ is a gauge connection
of the non-compact Lie group $SO(2,1)$, and so the kinetic terms in the Lorentz
invariant action for the minimal coupling of the spin-$j$ fields $\Phi^{(j)}$
to the spin-connection will be constructed from the usual gauge-covariant
derivatives
\beq
\nabla_\mu\Phi^{(j)}=\partial_\mu\Phi^{(j)}-i\Sigma_i^{(j)}\omega_\mu^i\Phi
^{(j)}
\label{omcovderivj}\eeq
where $\Sigma_i^{(j)}$ are the generators of the $(2j+1)$-dimensional spin-$j$
representation of $SO(2,1)$. To obtain a diffeomorphism invariant action, the
covariant derivatives (\ref{omcovderivj}) will appear contracted with the
metric tensor density $\sqrt{g}~g^{\mu\nu}$ of the spacetime which will lead to
additional interactions with the dreibein fields $e_\mu^i$. Since the
gravitational Chern-Simons term can be regarded as an $SO(2,1)$ gauge field
Chern-Simons term, we could naively expect that the induced spin from the
exchange of one spin-connection coincides with the tree-level gauge theory
result in (\ref{dimexpk}),(\ref{Delta1CS}). Then replacing $k$ by $k'$ in
(\ref{dimexpk}) and noting that the quadratic Casimir eigenvalue of the
spin-$j$ representation of $SO(2,1)$ is
\beq
T_j[SO(2,1)]=-j(j-1)
\label{TRSO21}\eeq
we see that this naive evaluation leads to the order $1/k'$ term in the KPZ
formula (\ref{KPZj}). However, this heuristic argument, which connects the
basic symmetry group of topologically massive gravity with the hidden
$SL(2,{\Bbb R})$ symmetry group of two-dimensional quantum gravity, is not
quite precise because the spin-connection is related to the dreibein field by
the Cartan-Maurer equation $\nabla e^a=0$. For this to be the total tree-level
conformal weight requires that the parity odd parts of the one-graviton and
$\omega e$ exchange interactions vanish. In the following we shall see that
this is indeed the case, i.e. the total tree-level conformal dimension,
described as above by a parity-odd Aharonov-Bohm type scattering amplitude, is
determined solely by the interaction with the spin-connection $\omega$. This
property is, as we will see, a consequence of the dynamical nature of the
gravitationally interacting particles \cite{a-cks}.

Notice though that this argument breaks down for higher-loop orders. For
instance, the one-loop (two-gluon) exchange contribution to the conformal
weight in Chern-Simons gauge theory coupled to charged scalar fields is
\cite{a-cks}
\beq
\Delta^{(2)}=-C_2[G]~{\rm sgn}(k)
\label{Delta2CS}\eeq
with $C_2[G]$ the dual Coxeter number of the gauge group $G$. The weight
(\ref{Delta2CS}) does not coincide with the ${\cal O}(1/k'^2)$ term in the
expansion (\ref{KPZj}) using the above naive arguments. We do expect that the
amplitudes corresponding to the exchange of two spin connection fields will
contribute a term analogous to (\ref{Delta2CS}) (with $C_2[SO(2,1)]=-2$), but
there will be additional non-vanishing contributions from amplitudes involving
a mixing of the spin connection with the other gravity fields. Furthermore, as
discussed in \cite{a-cks}, in the gravitational case we do not anticipate a
dependence on the sign of the Chern-Simons coefficient $k'$, as is the case in
Chern-Simons gauge theory. The contributions from sole graviton (and other
mediating boson) exchanges that do not involve $\omega$ do, however, vanish
\cite{a-cks}. The matter-coupled topologically massive gravity theory in this
way provides a dynamical illustration of the role of the $SL(2,{\Bbb R})\cong
SO(2,1)$ current algebra symmetry of quantum Liouville theory.

In the following we shall compute tree-level amplitudes in the topologically
massive gravity theory coupled to spinning fields. First, we point out two
features of the above identification of the conformal dimensions. The first one
is that the gravitational renormalization (\ref{KPZj}) produces only one chiral
component of the dressed spin. The full spin in the conformal field theory is
determined as the difference $\Delta-\bar\Delta$ of the weights for the
holomorphic and anti-holomorphic sectors of the world-sheet theory. The
Aharonov-Bohm amplitude therefore only describes the holomorphic (or
anti-holomorphic) observables in the induced conformal field theory on
$\partial{\cal M}$, i.e. the spins with $\bar\Delta=0$. Thus, strictly
speaking, the induced spins that we calculate in this way are really the chiral
scaling dimensions. This is evident from the relationship between vertex
operators of the induced two-dimensional conformal field theory on
$\partial{\cal M}$ and Wilson line operators of the three-dimensional gauge
theory in $\cal M$ that was discussed in Section 1.

The second feature concerns the structure of the tree-level amplitude for
fermion fields of mass $m$ in a representation $R[G]$ of the gauge group $G$
minimally coupled to a Chern-Simons gauge field. Then the matter field current
is $J^\mu_a=\bar\psi_A\gamma^\mu R_{AB}^a\psi_B$ (see the next Section), and
the scattering amplitude is \cite{dekss}
\beq
{\cal A}_f(p_1,p_2;q)=-\frac{4\pi\dim(G)T_R[G]}{km}+\frac{4\pi
i\dim(G)T_R[G]}{km^2}\frac{\epsilon_{\mu\nu\lambda}p_1^\mu p_2^\nu
q^\lambda}{q^2}
\label{AfCS}\eeq
in the infrared limit $q^2\to0$. The first term comes from the finite
renormalization of the $U(1)$ charge and corresponds to a short-ranged Pauli
magnetic moment interaction arising from the bare spin of the fermion fields.
The imaginary, parity-odd second term in (\ref{AfCS}) which has a simple pole
at $q^2=0$ leads to the usual Aharonov-Bohm interaction and identifies the
induced spin as in (\ref{Delta1CS}). Note that by dimensional analysis the
spinor vertices are down by a factor of $2m$ with respect to those of scalar
fields, so that the overall amplitude is suppressed by a factor of $4m^2$ in
its relation to the conformal dimensions.

\section{KPZ Weights as $SO(2,1)$ Anomalous Spin}

Given the above identification of the KPZ conformal weights as the parity-odd,
singular pole terms of the scattering amplitudes, we shall now demonstrate
explicitly how this structure appears in topologically massive gravity. The
relevant Aharonov-Bohm type contributions can come from the parity odd parts of
the gravitational propagators in Section 2, but, as we shall now discuss, there
are some subtleties in this description in the pure gravitational case. In this
Section we present explicit tree-level calculations to illustrate the
discussion above.

\subsection{Scalar Representations}

The simplest case of charged scalar fields $\Phi^{(0)}$ (i.e. the trivial spin
$j=0$ representation $\Sigma^{(0)}_i=0$) coupled to gravity is described by the
action
\beq
S_s=\int_{\cal M}\left((\nabla\phi)^*\wedge\star\nabla\phi-m^2\phi^*\star\phi
\right)=\int_{\cal
M}d^3x~\sqrt{g}~\left(g^{\mu\nu}\partial_\mu\phi^*\partial_\nu\phi-m^2~
\phi^*\phi\right)
\label{scalaraction}\eeq
where the meson mass $m$ is used to regulate infrared divergences arising from
the matter loops. In this case, there is no interaction with the
spin-connection and the parity-odd parts of the graviton exchange diagrams
vanish to all orders of perturbation theory \cite{a-cks}. To see this, we use
the shifts (\ref{dreishift}) to expand the metric determinant factor in
(\ref{scalaraction}) as
\beq
\sqrt{g}=1-\frac{1}{2}h^\mu_\mu+\frac{1}{4}\left(\frac{1}{2}h^\mu_\mu
h^\nu_\nu-h^\mu_\nu h^\nu_\mu\right)+\dots
\label{sqrtgexp}\eeq
where
\beq
g_{\mu\nu}=\eta_{\mu\nu}+e_\mu^ae_\nu^a-\eta_{\mu a}e_\nu^a-\eta_{\nu
a}e^a_\mu\equiv\eta_{\mu\nu}+h_{\mu\nu}
\label{metricexp}\eeq
is the expansion of the dynamical metric field about the flat background. Then
the meson-meson-graviton vertex is
\beq
{\cal E}_i^\mu(p,p';q)=i(\delta^\mu_i[p\cdot p'-m^2/2]+p'^\mu p_i+p^\mu p'_i)
\label{mmgvertex}\eeq
where $q=p-p'$, and the parity-odd $\epsilon$-part of the tree-level amplitude
for the exchange of one graviton between the charged mesons is (Fig. 1 with the
straight lines denoting the meson fields and the wavy line representing the
dreibein field $e_\mu^a$)
\beq
{\cal B}_s(p_1,p_2;q)^{\rm odd}=i{\cal E}_i^\mu(p_1,p_1-q){\cal
E}_j^\nu(p_2,p_2+q)D_{\mu\nu}^{ij}(q)^{\rm odd}\equiv0
\label{treegrav}\eeq
The tree amplitude (\ref{treegrav}) vanishes exactly for all ranges of the
momentum transfer $q$. The vanishing at higher-loop orders is then a general
consequence of the minimal coupling of the dreibein fields in a transverse
gauge to the conserved, dynamical gravitational particle current \cite{a-cks}.

Thus, in the case of the coupling of topologically massive gravity to dynamical
charged scalar fields, we can reproduce correctly the KPZ formula (\ref{KPZj})
for a bare spin $j=0$, i.e.
\beq
\Delta_0(k')=0
\label{0spin}\eeq
It was also shown in \cite{a-cks}, up to one-loop order, that the gravitational
dressing of an anomalous fractional spin $\Delta_0=1/k$ of the scalar fields
also leads to the anticipated result. In the following we shall be concerned
with this renormalization for higher-spin representations of the local Lorentz
group of the spacetime.

\subsection{Spinor Representations}

Next we consider the coupling of topologically massive gravity to fields
$\Phi^{(1/2)}$ in the lowest non-zero spin $j=1/2$ representation of the local
Lorentz group. The coupling to spinor fields in the (2 + 1)-dimensional case is
especially intriguing because the gamma-matrices obey the identity
\beq
\gamma_\mu\gamma_\nu=g_{\mu\nu}+i\epsilon_{\mu\nu\lambda}\gamma^\lambda
\label{2+1gammaid}\eeq
and so the Dirac matrices themselves generate a representation of the $SO(2,1)$
Lie algebra, i.e. $\Sigma_i^{(1/2)}=\gamma_i/2$. The irreducible
two-dimensional spinor representation is defined by the Pauli spin matrices
\beq
\gamma_0=\sigma^3=\pmatrix{1&0\cr0&-1\cr}~~~,~~~\gamma_1=i\sigma^1=\pmatrix{
0&i\cr i&0\cr}~~~,~~~\gamma_2=i\sigma^2=\pmatrix{0&1\cr-1&0\cr}
\label{spingamma}\eeq
The unique, Lorentz-invariant and lowest order derivative action for Dirac
fermion fields is defined by the usual Dirac kinetic term with a minimal
coupling to the spin-connection $\omega_\mu^i$ as described above. Thus we can
couple topologically massive gravity to fermion fields by the action
\beq
S_F=\int_{\cal M}\left(\bar\psi~i\gamma\wedge\star\nabla\psi-m\bar\psi\star\psi
\right)=\int_{\cal M}d^3x~\left[\sqrt{g}~g^{\mu\nu}\bar\psi~i\gamma_\mu\left(
\partial_\nu-\frac{i}{2}\gamma_i~\omega^i_\nu\right)\psi-\sqrt{g}~m
\bar\psi\psi\right]
\label{SFdef}\eeq
where $\psi(x)$ are two-component fermion fields and $m$ is the fermion mass
which is again introduced to avoid infrared divergence problems. Note that in a
three-dimensional spacetime $m$ can be either positive or negative. To find the
tree level gravity interactions, we use the metric expansions (\ref{sqrtgexp})
and (\ref{metricexp}). For the tree-level amplitudes, it suffices to keep only
terms linear in the dreibein field. Then the $\bar\psi\psi\omega$ vertex is
\beq
{\cal
W}^\mu_i(p,p';q)=-\frac{i}{2}\gamma^\mu\gamma_i=-\frac{i}{2}\left(\delta^\mu_i
+i\epsilon^\mu_{~i\lambda}\gamma^\lambda\right)
\label{ffomega}\eeq
where $q=p-p'$ and we have used (\ref{2+1gammaid}). The $\bar\psi\psi e$ vertex
is
\beq
{\cal G}_i^\mu(p,p';q)=\frac{i}{2}\left[2(p+p')^\mu\gamma_i+\delta_i^\mu\gamma^
\lambda(p+p')_\lambda+2m\delta^\mu_i\right]
\label{ffe}\eeq
with $q=p-p'$.

There are four tree-level amplitudes in this gravitationally dressed field
theory, which are represented in Fig. 1 with the straight lines denoting the
fermion fields and the wavy line depicting an exchange of the spin-connection,
graviton, $\omega e$ or $e\omega$ fields. First, we consider the
renormalization due to the spin-connection. The amplitude is
\beq
{\cal A}_f^{({\rm grav})}(p_1,p_2;q)=\frac{4\pi i}{k'}\bar u(p_1-q){\cal
W}^\mu_iu(p_1)~\Omega^{ij}_{\mu\nu}(q)~\bar u(p_2+q){\cal W}^\nu_ju(p_2)
\label{AfTMG}\eeq
where $u(p)$ are the on-shell positive energy Dirac spinors. They obey the
momentum space Dirac equation
\beq
(p_\mu\gamma^\mu-m)u(p)=\bar u(p)(p_\mu\gamma^\mu-m)=0
\label{Diraceq}\eeq
whose solutions with the Bjorken-Drell normalization $\bar u(p)u(p)=1$ are
\beq
u(p)=\frac{1}{2m(p^0+m)}\pmatrix{p^0+m\cr-i(p^1+ip^2)\cr}
\label{upsolns}\eeq
with the on-shell condition $(p)^2=(p^0)^2-(p^1)^2-(p^2)^2=m^2$. Here we have
assumed for definiteness that $m>0$.

Using (\ref{ffomega}) along with the propagator identities (\ref{spinpropids})
listed in the Appendix at the end of the Paper, we find
\beq\new{\begin{array}{lll}
{\cal A}_f^{({\rm
grav})}(p_1,p_2;q)&=&-\frac{\pi}{2k'M}\left\{\left(3-\frac{q^2}{M^2}
\right)\bar u(p_1-q)u(p_1)\bar u(p_2+q)u(p_2)\right.\\&
&~~~~~+\frac{2(q^2+M^2)}{M(q^2-M^2)}\left(q_\lambda\bar u(p_2+q)\gamma^\lambda
u(p_2)\bar u(p_1-q)u(p_1)\right.\\& &\left.~~~~~-q_\lambda\bar
u(p_1-q)\gamma^\lambda u(p_1)\bar u(p_2+q)u(p_2)\right)\\&
&~~~~~\left.+2M\Omega_{\mu\nu}^{ij}(q)\epsilon^i_
{~\mu\lambda}\epsilon^j_{~\nu\rho}\bar u(p_1-q)\gamma^\lambda u(p_1)\bar
u(p_2+q)\gamma^\rho u(p_2)\right\}\end{array}}
\label{ffamplsimpl}\eeq
The amplitude (\ref{ffamplsimpl}) can be simplified using the Dirac equation
(\ref{Diraceq}) and the identity (\ref{spinpropepid}). We are interested in the
small momentum limit of this amplitude, i.e. $q^2\to0$. In that limit, the
fermion bilinears appearing above become
\beq
\bar u(p\pm q)u(p)\to1~~~~~,~~~~~\bar u(p\pm q)\gamma^\mu u(p)\to(2p\pm
q)^\mu/2m
\label{spinorsq0}\eeq
and we arrive at
\beq
{\cal A}_f^{({\rm grav})}(p_1,p_2;q)=\frac{\pi}{2k'M}+\frac{3\pi
i}{k'm^2}\frac{\epsilon_{\mu\nu\lambda}p_1^\mu p_2^\nu q^\lambda}{M^2}
\label{Affinal}\eeq

The imaginary, parity-odd part of the amplitude (\ref{Affinal}) does not have a
singular pole term at $q^2=0$ like in the gauge theory case, so it does not
really represent a genuine Aharonov-Bohm interaction amplitude. The entire
amplitude vanishes in the infrared limit $M\to\infty$ when the topologically
massive gravity theory becomes topological Einstein gravity. This is similar to
the effect of taking $k\to\infty$ in the Chern-Simons gauge amplitude
(\ref{AfCS}). The ``pole" structure here appears as $M^2\to0$ (equivalently
$\kappa\to0$) when the gravitational Chern-Simons action in (\ref{TMGaction})
dominates. In fact, the zero-mass limit of topologically massive gravity is
equivalent to pure topological $SO(2,1)$ Chern-Simons gauge theory
\cite{kogan1}. In this regime, where $M^2\to0$ with $q^2\ll M^2$, the moduli
space structure of the induced two-dimensional gravity theory disappears. The
value $\Lambda=0$ of the boundary cosmological constant corresponds to a
critical point of the three-dimensional quantum field theory, namely the point
where the topologically massive gravity action exhibits local (2 +
1)-dimensional conformal invariance \cite{djt}. The induced spin can be
identified from (\ref{Affinal}) as a quantity which is independent of the
topological graviton mass scale, so that the singularity at the conformal
symmetry point $M^2=0$ can be regarded as a kinematical pole corresponding to
the usual non-dynamical singularity characteristic of a pure Chern-Simons gauge
amplitude (with gauge group $SO(2,1)$). We can use these facts to interpret the
imaginary part of (\ref{Affinal}) as an Aharonov-Bohm type amplitude, and
comparing with (\ref{AfCS}) this identifies the gravitationally-dressed
spin-$\frac{1}{2}$ weight
\beq
\Delta_{1/2}^{(1)}=1/4
\label{1/2weight}\eeq
which agrees with the coefficient of the ${\cal O}(1/k')$ term in (\ref{KPZj})
for $j=1/2$. Later on we will describe another way of determining the anomalous
conformal dimension from the parity-odd structure in (\ref{Affinal}).

The feature that the gravitational Chern-Simons term does not induce a singular
momentum pole term owes to the fact that the quadratic form for the
spin-connection in the topologically massive gravity action (\ref{TMGshift}) is
non-degenerate. This is because the $\omega\wedge de$ term in the Einstein
action acts like a Proca mass term for a gauge field, thus removing all
infrared singular pole terms. It does, however, induce a parity-odd structure
indicative of the renormalization of the spin, where we naively interchange the
roles $M^2\leftrightarrow q^2$ for the above correspondences. With this
interchange the low-energy structure of topologically massive gravity begins to
resemble that of topologically massive gauge theory \cite{djt}, i.e. one with
both a Yang-Mills kinetic term and a topological Chern-Simons term for the
gauge fields. The structure of these amplitudes in topologically massive
gravity is opposite to those of the gauge theory. This is another indication of
the sort of duality that has been previously observed in \cite{a-cks} for the
perturbative, infrared structure of these two topologically massive quantum
field theories.

To explore this $q^2=0$ behaviour further, let us now consider the one-graviton
exchange amplitude in the infrared limit
\beq\new{\begin{array}{lll}
{\cal B}_f(p_1,p_2;q)&=&i\bar u(p_1-q){\cal
G}^\mu_i(p_1,p_1-q)u(p_1)~D_{\mu\nu}^{ij}(q)~\bar u(p_2+q){\cal
G}^\nu_j(p_2,p_2+q)u(p_2)\\&=&-i\left[4\bar u(p_1-q)\gamma^iu(p_1)\bar
u(p_2+q)\gamma^ju(p_2)p_1^\mu p_2^\nu
D_{\mu\nu}^{ij}(q)+4m^2D_{ij}^{ij}(q)\right.\\& &~~~~~~\left.+2m\bar
u(p_1-q)\gamma^iu(p_1)p_1^\mu D_{\mu j}^{ij}(q)+2m\bar
u(p_2+q)\gamma^ju(p_2)p_2^\nu D_{i\nu}^{ij}(q)\right]\end{array}}
\label{BfTMG2}\eeq
where we have used the on-shell conditions $p^2=(p\pm q)^2=m^2$, the Dirac
equation (\ref{Diraceq}), and transversality of the free graviton propagator.
In the small $q^2$ limit, using the propagator identities (\ref{gravpropids})
listed in the Appendix we find
\beq
{\cal B}_f(p_1,p_2;q)=\frac{60\pi m^2(q^2+M^2)}{k'Mq^2(q^2-M^2)}
\label{Bffinal}\eeq
Thus the one-graviton exchange amplitude has only a parity-even part, and so it
does not contribute to the anomalous spin of the fields. Furthermore, it has a
singular pole structure at $q^2=0$, reflecting the fact that the original
Einstein theory ($M\to\infty$ in (\ref{TMGaction})) can be regarded as a
topological Chern-Simons theory with gauge group $ISO(2,1)$. This topological
Einstein theory contains no propagating degrees of freedom, which is why the
characteristic pole at $q^2=0$ appears here, as in the case of the topological
pure Chern-Simons gauge theory.

Finally, we consider the amplitude for the exchange of one $e\omega$ boson,
\beq
{\cal C}_f(p_1,p_2;q)=i\bar u(p_1-q){\cal W}_i^\mu
u(p_1)~E_{\mu\nu}^{ij}(q)~\bar u(p_2+q){\cal G}_j^\nu(p_2,p_2+q)u(p_2)
\label{Cftree}\eeq
Using the spin-connection propagator identities (\ref{spinpropids}),
(\ref{qspinpropid}) and (\ref{spinpropepid}) we find that the $q^2\to0$ limit
of the amplitude (\ref{Cftree}) becomes
\beq
{\cal C}_f(p_1,p_2;q)=\frac{3\pi}{k'M}\left(3+\frac{4m}M\right)+\frac{4\pi
i}{k'm^2}\frac{\epsilon_{\mu\nu\lambda}p_1^\mu p_2^\nu
q^\lambda}{M^2}\left(\frac{q^2}{M^2}\right)
\label{Cfq20}\eeq
The parity-odd term and the term proportional to the fermion mass $m$ in the
parity-even part of the amplitude (\ref{Cfq20}) come from the spin-connection
part of the $e\omega$ propagator $E_{\mu\nu}^{ij}(q)$ in (\ref{gravprop}),
while the remaining parity-even piece comes from the $\beta e$ part of
$E_{\mu\nu}^{ij}(q)$. Although the parity-odd piece in (\ref{Cfq20}) has a
similar structure as that which appears in the $\omega$-exchange amplitude, it
is of order $q^2/M^2$ and thus vanishes in the infrared regime $q^2\ll M^2$ as
compared to the amplitude (\ref{Affinal}), i.e. it vanishes in the correlated
limit above in which we take first take $q^2\to0$ and then identify the
kinematical-type pole at $M^2\to0$. This results from the manner in which the
extra spin-connection term is convoluted in (\ref{Ddef}) with the spin matrix
$\Sigma^{kl}$ to yield the transverse projection of $\Omega_{\mu\nu}^{ij}(q)$
in (\ref{gravprop}). Thus the $\omega e$-exchange amplitude leads to ${\cal
O}(q^2/M^2)$ corrections to the parity-odd, Aharonov-Bohm type amplitude
determining the conformal weights and so, by definition, it yields no
contribution to the induced spin at tree-level in perturbation theory. It would
be interesting to further explore the (relativistic) structure of the
parity-odd amplitudes above more carefully and hence study more precisely the
interactions between topologically massive gravity and spinning matter fields,
and also its potential relevence to the genuine gravitational Aharonov-Bohm
effect \cite{desmc,a-cksgrav}.

The above structures that are a result of the form of the gravitational
propagators will play an important role in the higher-loop scattering
amplitudes. The parity-odd structures from the graviton and $e\omega$ lines
will always vanish at $q^2\to0$, which is a result of the index contractions
from the vertices formed by the kinetic terms of the matter fields coupled to
the gravity fields \cite{a-cks}. These parity-odd pieces, which are a result of
the covariant derivative relation between the $e$ and $\omega$ fields, are
expected to vanish in order to recover the pure Einstein theory results.
However, in diagrams with {\it both} graviton or $e\omega$ and spin-connection
lines, the parity-odd part of the spin-connection propagator will combine with
the parity-even, singular part of the graviton or $e\omega$ propagators, thus
producing the required Aharonov-Bohm interaction terms. It is this sort of
interplay between the $e$ and $\omega$ fields that will lead to a more
complicated conformal dimension in higher loops than that anticipated from a
naive gauge theory calculation. This sort of renormalization at higher-loop
orders is exemplified in the gravitational dressing of an induced spin from the
interaction of charged scalar fields with ordinary Chern-Simons gauge theory
\cite{a-cks}. The parity-even parts of the graviton propagator
$D_{\mu\nu}^{ij}(q)$ renormalize the parity-odd part of the Chern-Simons gluon
propagator and conspire to yield the anticipated KPZ scaling at one-loop order
(in the Einstein limit $M\to\infty$). The remaining combinations, however,
vanish identically. The fact that such an interplay doesn't appear at
tree-level here is the reason why there is no singular pole term in the
$\omega$-exchange amplitude, but it still does allow the identification of the
correct KPZ induced spin as above. It would be interesting to extend this
calculation to one-loop order and match the diagrammatic contributions with the
${\cal O}(1/k'^2)$ term in the KPZ formula. It would also be interesting to see
how these higher-loop amplitudes behave in the two limits $M\to\infty$ and
$M\to0$ where topologically massive gravity is equivalent to topological
Chern-Simons gauge theories (with gauge groups $ISO(2,1)$ and $SO(2,1)$,
respectively).

\subsection{Vector Representations}

Finally, we couple to fields $\Phi^{(1)}$ in the defining, vector
representation of $SO(2,1)$. The simplest Lorentz- and gauge-invariant model is
the abelian topologically massive gauge theory
\beq
S_V=\int_{\cal M}-\frac1{4e^2}(\nabla A)^*\wedge\star\nabla
A+S_{CS}^{[U(1)]}=\int_{\cal
M}d^3x~\left(-\frac{1}{4e^2}~\sqrt{g}~g^{\mu\lambda}g^{\nu\rho}F_{\mu\nu}
F_{\lambda\rho}+\frac{k}{8\pi}\epsilon^{\mu\nu\lambda}A_\mu\partial_\nu
A_\lambda\right)
\label{SVdef}\eeq
The Chern-Simons term is introduced to regulate the logarithmic infrared
divergences in the pure Maxwell theory by giving the photons of the model a
topological mass $m=ke^2/4\pi$. Since this term is topological, it does not
couple to the spacetime metric and so all results below will be valid even in
its absence. Now there is no coupling to the (torsion-free) connection $\omega$
because the spin-connection terms in the covariant derivatives in
(\ref{omcovderivj}) cancel out by anti-symmetry in the curvature
$F_{\mu\nu}=\nabla_\mu A_\nu-\nabla_\nu A_\mu=\partial_\mu A_\nu-\partial_\nu
A_\mu$. Thus the contribution to the conformal weight from the $\omega$ and
$\omega e$ exchange amplitudes as described above is 0.

It remains to check if there is any contribution from a parity-odd structure of
the one-graviton exchange amplitude. Shifting the dreibein fields as above, the
graviton-photon-photon vertex is \cite{a-cks}
\beq
{\cal F}^{\mu}_{i;\nu\rho}(p;q,r)=\frac{i}{2e^2}\left[(r\cdot
p)\delta^\mu_i\eta_{\nu\rho}-\delta^\mu_ir_\nu p_\rho+p^\mu
r_\nu\eta_{i\rho}+2r_ip_\rho\delta^\mu_\nu-2(r\cdot p)\delta^\mu_\nu
\eta_{i\rho}-2p^\mu r_i\eta_{\nu\rho}\right]
\label{gppvertex}\eeq
where $p=q+r$, which leads to the exchange amplitude (Fig. 1 with the straight
lines representing the photon fields and the wavy line depicting the dreibein
field)
\beq
{\cal B}_v(p_1,p_2;q)=ie_\nu^*(p_1-q){\cal
F}^\alpha_{i;\mu\nu}(p_1,p_1-q)e_\mu(p_1)~D_{\alpha\beta}^{ij}(q)~
e_\rho^*(p_2+q){\cal F}^\beta_{j;\lambda\rho}(p_2,p_2+q)e_\lambda(p_2)
\label{BvTMG}\eeq
where $e(p)$ are the on-shell polarization vectors for the photons in the
transverse Landau gauge. These vectors describe the propagation of the single
gauge degree of freedom in the quantum field theory (\ref{SVdef}) and are
determined from the equations of motion
\beq
(\eta_{\mu\nu}\Box+m\epsilon_{\mu\nu\lambda}\partial^\lambda)A^\nu=0
\label{coveqmotion}\eeq
in a covariant gauge $\partial_\mu A^\mu=0$. In momentum space,
(\ref{coveqmotion}) leads to the on-shell equation
\beq
me_\mu(p)=-i\epsilon_{\mu\nu\lambda}p^\nu e^\lambda(p)
\label{momeqmotion}\eeq
which has solution
\beq
e(p)=\frac{1}{\sqrt{2}m|\vec p|}\pmatrix{\vec p^2\cr p^0p^1-imp^2\cr
p^0p^2+imp^1\cr}
\label{polvecs}\eeq
where $(p)^2=(p^0)^2-\vec p^2=m^2$, and we have used the transverse
normalizations
\beq
p^\mu e_\mu(p)=0=e^\mu(p)e_\mu(p)~~~~~,~~~~~e^\mu(p)e_\mu^*(p)=-1
\label{polids}\eeq

In the limit $q^2\to0$, using (\ref{polvecs}) and (\ref{polids}) we can
simplify the amplitude (\ref{BvTMG}) to
\beq\new{\begin{array}{lll}
{\cal B}_v(p_1,p_2;q)&=&-\frac{i(q^2)^2}{16e^4}\left[2e_i^*(p_1-q)e^\alpha
(p_1)D^{ij}_{\alpha j}(q)+2e_j^*(p_2+q)e^\beta(p_2)D^{ij}_{i\beta}(q)\right.\\&
&~~~~~~~~~~\left.+4e_i^*(p_1-q)e_j^*(p_2+q)e^\alpha(p_1)e^\beta(p_2)
D^{ij}_{\alpha\beta}(q)+D^{ij}_{ij}(q)\right]\end{array}}
\label{Bv1}\eeq
Using (\ref{polvecs}) we have
\beq
\epsilon_{\mu~\lambda}^{~j}e^\mu(p)e_j^*(p\pm q)q^\lambda=-\frac{ip\cdot
q}{m}=\mp\frac{iq^2}{2m}
\label{polepsid}\eeq
where $p=p_1$ or $p_2$, respectively. Using other similar such identities
derived from (\ref{polvecs}) and (\ref{gravpropids}), we find that all
parity-odd contributions in (\ref{Bv1}) vanish in the infrared regime. The
total amplitude in the limit $q^2\to0$ is then
\beq
{\cal B}_v(p_1,p_2;q)=\frac{\pi}{2k'e^4M}~q^2
\label{Bvfinal}\eeq
where the usual pole at $q^2=0$ has been cancelled in this case by the
higher-momentum interactions from the vertex function (\ref{gppvertex}). Thus
the gravitational renormalization of the scaling dimension of the spin-1 fields
here is
\beq
\Delta_1^{(1)}=0
\label{spin1weight}\eeq
which again agrees with the coefficient of the ${\cal O}(1/k')$ term in the KPZ
formula (\ref{KPZj}) for spin $j=1$.

We stress that the qualitative features of the above tree-level calculations
are valid for any range of momentum $q$. The parity-odd parts of the graviton
exchange amplitudes always vanish, because of the contractions of the free
graviton propagator $D_{\mu\nu}^{ij}(q)$ with a conserved current, while the
amplitudes involving the spin-connection lead to the required parity-odd
structure identifying the conformal dimensions. These latter amplitudes do not
contain the kinematic zero-momentum pole characteristic of Aharonov-Bohm
scattering. However, the $q^2\to0$ limits of all of these amplitudes are
independent of the mass scale of the theory (modulo the $1/M^2$ behaviour of
the parity-odd parts), i.e. they hold independently of the size of the mass
ratio $m/M$. It is this scale independence that is the remarkable feature of
the (2 + 1)-dimensional gravitational dressing of the spins. This feature was
also pointed out in \cite{a-cks}. In the next Section we shall present an
alternative way of describing the spectrum of anomalous dimensions in
topologically massive gravity.

\section{KPZ Weights as $ISO(2,1)$ Anomalous Spin}

In the previous Section we demonstrated the role of the local $SO(2,1)$
spacetime symmetry group of topologically massive gravity in its relation to
the quantum Liouville theory. It illustrates explicitly how the mysterious
``hidden" $SL(2,{\Bbb R})$ Kac-Moody symmetry of two-dimensional quantum
gravity \cite{kpz} arises as a dynamical property of the coupling of
three-dimensional gravity to sources. This realization of the
gravitationally-dressed scaling weights is completely independent of the mass
scale of the theory, but it requires tuning the topological graviton mass to a
neighbourhood of the conformally-invariant point in the parameter space of the
three-dimensional quantum field theory. This limit is required to properly
identify the Aharonov-Bohm type amplitudes in the low-energy regime of the
theory. The resulting scaling dimensions then appear as the usual induced
fluxes responsible for the Aharonov-Bohm phase factors characteristic of the
reduction of topologically massive gravity to topological $SO(2,1)$
Chern-Simons gauge theory.

The unusual limiting procedure required above suggests that we should look for
an alternative way to identify the anomalous conformal dimensions which does
not involve taking an unusual correlated low-energy limit as above. For this,
we exploit the fact that we are really interpreting the scaling weights as the
induced spin of the matter fields from their interaction with topologically
massive gravity and consider the explicit renormalization of spin. The
parity-odd structure of topologically massive gravity allows a very natural way
of doing this in (2 + 1)-dimensions. As we shall see, this leads to a
renormalized weight which is scale dependent and reduces to the anticipated KPZ
scaling weights in the other topological limit of topologically massive
gravity, namely the limit of pure Einstein gravity (equivalently $ISO(2,1)$
Chern-Simons gauge theory). In this way we will have a dynamical description of
the scaling dimensions in the full parameter regime of topologically massive
gravity which incorporates both of the fundamental symmetry groups $SO(2,1)$
and $ISO(2,1)$ of the three-dimensional quantum field theory, as well as the
discrete parity-violating property of Chern-Simons quantum field theory. This
illustrates the potential relevance of the geometric symmetries of the
topological membrane approach to the gravity sector of string theory.

\subsection{Anomalous Magnetic Moment in Topologically Massive Gravity}

As discussed in \cite{kogsem}, the geometry of a (2 + 1)-dimensional spacetime
allows the identification of the induced fractional spin in an unambiguous way
through the magnetic moment. The key feature is that the (2 + 1)-dimensional
gamma-matrix identity (\ref{2+1gammaid}) implies that the spin matrix generator
on the space of spin-$\frac12$ fields is given by
\beq
\sigma^{\mu\nu}\equiv\frac
i2\left[\gamma^\mu,\gamma^\nu\right]=-\epsilon^{\mu\nu\lambda}\gamma_\lambda
\label{spin12matrix}\eeq
and so the Gordon decomposition for the (2 + 1)-dimensional spinor current is
\beq
\bar u(p-q)\gamma^\mu u(p)=\frac{(2p-q)^\mu}{2m}\bar u(p-q)u(p)-\frac
i{2m}\epsilon^{\mu\nu\lambda}q_\lambda\bar u(p-q)\gamma_\nu u(p)
\label{gordondecomp}\eeq
In the non-relativistic limit $q^2\to0$ of the corresponding Dirac Hamiltonian,
the first term in (\ref{gordondecomp}) represents the Coulomb charge
interaction when the fermion fields are minimally coupled to an electromagnetic
vector potential $A_\mu$ (see (\ref{spinorsq0})). The other term represents the
Pauli magnetic moment interaction in the usual way. In the non-relativistic
limit, we see then that the quantity
\beq
\varrho^\mu=-\frac i{2m}\frac{\epsilon^{\mu\nu\lambda}p_\nu q_\lambda}m
\label{magmomdefrho}\eeq
describes the usual magnetic moment structure in the Pauli interaction.

The function $\varrho^\mu$ is an axial vector, and is therefore odd under
parity. The appearence of the magnetic moment structure in (2 + 1)-dimensions
is therefore very natural for a parity-violating quantum field theory, such as
Chern-Simons theory. It can be determined by examining the parity-odd structure
of the one-particle irreducible vertex function $\Gamma_\mu(p,q)$ which is
defined through the complete vertex function
\beq
V_\mu(p,q)\equiv\langle\psi(p-q)A_\mu(q)\bar\psi(p)\rangle=S(p-q)~i\Gamma_\mu
(p,q)~S(p)
\label{vertexfndef}\eeq
for the interaction of fermion fields minimally coupled to a background $U(1)$
gauge field $A_\mu$. Here $S(p)=\langle\psi(p)\bar\psi(-p)\rangle$ is the
renormalized fermion propagator in momentum space. The identity
(\ref{spin12matrix}) shows that the vertex function $\Gamma_\mu(p,q)$ can be
decomposed into parity-even and parity-odd pieces which define the usual form
factors by
\beq
\Gamma_\mu(p,q)=\gamma_\mu~\Gamma_e(q^2)+\frac
i{2m}\sigma_{\mu\nu}q^\nu~\Gamma_o(q^2)
\label{formfacdef}\eeq
The decomposition (\ref{formfacdef}) holds in the gravitationally-dressed
theory because of the discrete symmetries of the topologically massive gravity
action. To actually compute the irreducible vertex function (\ref{formfacdef})
we consider $\Gamma_\mu(p,q)$ to be contracted between Dirac spinors as $\bar
u(p-q)\Gamma_\mu(p,q)u(p)$. We shall not write the spinors explicitly but we
freely use the Gordon relation (\ref{gordondecomp}). It implies that the
renormalization of the magnetic moment structure (\ref{magmomdefrho}) is
determined by the low-energy limit of the sum $\Gamma_e(q^2)+\Gamma_o(q^2)$.
However, the longitudinal form factor $\Gamma_e(q^2)$ can be absorbed into
other renormalizations of the quantum field theory. This follows from the
Ward-Takahashi identity
\beq
q^\mu\Gamma_\mu(p,q)=\Sigma(p-q)-\Sigma(p)
\label{wardid}\eeq
where $\Sigma(p)$ is the fermion self-energy operator. As proven in
\cite{a-cks}, the standard set of Ward-Takahashi identities holds individually
for the radiative corrections due to gravity, because the gravity fields are
themselves coupled to conserved matter currents in the gravitational
interaction terms. The relation (\ref{wardid}) shows that the longitudinal
component of the vertex function (\ref{formfacdef}) can be cancelled by the
corresponding counterterms which renormalize the fermion mass, in addition to
the $U(1)$ fermion charge.

Thus the only contribution from the vertex function to the magnetic moment
structure $\varrho^\mu$ is effectively from the parity-odd form factor
$\Gamma_o(q^2)$, and so we can identify the magnetic moment $\mu$ of the
fermion fields as
\beq
\mu\equiv\frac{g}{2m}\Delta=-\frac12\lim_{q^2\to0}\Gamma_o(q^2)
\label{magmomdef}\eeq
where we have explicitly incorporated the cancellation of the self-energy
corrections to the external fermion lines with the parity-even component of the
irreducible vertex function. Here we assume that the initial and final fermions
are on-shell, $p^2=(p-q)^2=m^2$, and that they have gyromagnetic ratio $g=2$.
The magnetic moment (\ref{magmomdef}) can be used as an alternative definition
of the renormalized (transmuted) spin $\Delta$. Note that in the infrared limit
$q^2\to0$, we can make the replacements (\ref{spinorsq0}) and compute the
parity-odd form factor from
\beq
\Gamma_o(q^2)=\frac{8im}{q^2(q^2-4m^2)}\epsilon^{\mu\nu\lambda}p_\nu
q_\lambda\Gamma_\mu(p,q)
\label{gamodd}\eeq

Thus we can use the above discussion to obtain an unambiguous definition of the
renormalized spin $\Delta_{1/2}(k')$ for the interaction of fermion fields with
topologically massive gravity. For this, we add to our previous
gravitationally-dressed fermion action the term
\beq
S_A=\int_{\cal M}d^3x~\sqrt{g}~g^{\mu\nu}\bar\psi\gamma_\mu A_\nu\psi
\label{fermAgrav}\eeq
where $A$ is a non-dynamical photon field. Then we compute the irreducible
vertex function $\Gamma_\mu(p,q)$ for the renormalizations due to the
gravitational dressing order by order in $1/k'$. Defining the form factors as
above and using the relation (\ref{magmomdef}), we can then identify the
renormalized spin coefficients $\Delta^{(n)}_{1/2}$ and compare them with the
iterative expansion of the KPZ formula. This gives a very natural definition of
this induced spin for the spinor-coupled topologically massive gravity theory
which exploits the explicit parity-odd structure characteristic of the usual
identifications of the conformal dimensions. Furthermore, this alternative
definition of the scaling weights avoids the unusual limiting procedures of the
previous Section required to identify the spins using Aharonov-Bohm amplitudes.

\subsection{One-loop Proper Vertex Function}

As an explicit example of the above approach, we shall now compute the total
one-loop renormalization due to topologically massive gravity of the
one-particle irreducible fermion-fermion-photon vertex function. We use the
usual expansions of the metric about the flat background to obtain the relevant
vertices involving the background photon field at one-loop order. The
fermion-fermion-photon vertex is $-i\gamma_\mu$, the $\bar\psi\psi Ae$ vertex
is
\beq
{\cal H}^i_{\nu;\mu}(p,p';q;r)=-i(\gamma_\mu\delta^i_\nu+\gamma^i\eta_{\mu\nu}+
\gamma_\nu\delta_\mu^i)
\label{ffAe}\eeq
where $q+r=p-p'$, and the $\bar\psi\psi Aee$ vertex function is
\beq
{\cal H}^{ij}_{\nu\lambda;\mu}(p,p';q;r,s)=-i\left(\gamma^i\delta^j_\mu
\eta_{\nu\lambda}+\gamma^j\delta_\nu^i\eta_{\mu\lambda}+\gamma^i
\delta_\lambda^j\eta_{\mu\nu}-\frac12\gamma_\mu\eta_{\nu\lambda}\eta^{ij}+
\frac12\gamma_\mu
\delta^i_\nu\delta_\lambda^j-\gamma_\mu\delta^i_\lambda\delta_\nu^j\right)
\label{ffAee}\eeq
with $q+r+s=p-p'$.

The nine diagrams which contribute to the one-loop order renormalization of the
irreducible vertex function $\Gamma_\mu(p,q)$ are shown in Fig. 2. The first
diagram is the triangle graph from the exchange of the spin-connection field
and it will be evaluated explicitly below. The next three Feynman graphs
involving exchanges of the dreibein field do not contribute to the parity-odd
part of the vertex function. This can be checked explicitly using the above
Feynman rules and those in Section 4.2, and it follows from the general feature
that this parity-odd structure is always cancelled from the coupling of the
transverse dreibein fields to the conserved, dynamical gravitational matter
currents \cite{a-cks}. The fifth diagram in Fig. 2 is the graviton tadpole
graph and it is given explicitly by
\beq
\Gamma_\mu^{[\bigcirc]}(p,q)=\int\frac{d^3k}{(2\pi)^3}~{\cal
H}^{ij}_{\nu\lambda;\mu}D_{ij}^{\nu\lambda}(k)
\label{tadpole}\eeq
After some algebra the parity-odd part of (\ref{tadpole}) is readily seen to
vanish,
\beq
\Gamma_\mu^{[\bigcirc]}(p,q)^{\rm odd}=\frac{16\pi
i}{k'}\epsilon_{\mu\nu\lambda}\gamma^\lambda\int\frac{d^3k}{(2\pi)^3}~
\frac{k^\nu}{k^2(k^2-M^2)}\equiv0
\label{tad0}\eeq
after an appropriate dimensional regularization of the Feynman integral. The
last four diagrams involving exchanges of the $\omega e$ fields can generally
have finite contributions to $\Gamma_o(q^2)$ for non-zero $q^2$ due to the
spin-connection propagator. However, because of the form of the coupling to the
transverse graviton propagator and the transverse projection of
$\Omega_{\mu\nu}^{ij}(q)$ (see Section 4.2), these contributions will vanish in
the non-relativistic limit $q^2\to0$.  Again this is a feature of the coupling
of the transverse gravitational fields to the conserved matter currents
\cite{a-cks}.

\begin{figure}
\begin{center}
\begin{picture}(40000,7500)
\small
\drawline\fermion[\E\REG](0,5000)[10000]
\multiput(5000,4800)(0,-400){12}{\line(0,300){100}}
\drawline\photon[\NE\REG](1900,5000)[5]
\drawline\photon[\NW\FLIPPED](8200,5000)[5]
\put(4000,500){\makebox(0,0){$q$}}
\put(6000,500){\makebox(0,0){$\mu$}}
\put(0,6000){\makebox(0,0){$p$}}
\put(10000,6000){\makebox(0,0){$p-q$}}
\drawline\fermion[\E\REG](14000,5000)[10000]
\drawline\gluon[\NE\REG](16100,5000)[2]
\drawline\gluon[\NW\FLIPPED](21900,5000)[2]
\multiput(19000,4800)(0,-400){12}{\line(0,300){100}}
\drawline\fermion[\E\REG](28000,5000)[10000]
\drawline\gluon[\NW\FLIPPED](34300,5000)[1]
\drawline\gluon[\NE\REG](30500,5000)[1]
\multiput(30500,4800)(0,-400){12}{\line(0,300){100}}
\end{picture}
\end{center}
\begin{center}
\begin{picture}(40000,10000)
\small
\drawline\fermion[\E\REG](0,5000)[10000]
\drawline\gluon[\NW\FLIPPED](7500,5000)[1]
\drawline\gluon[\NE\REG](3700,5000)[1]
\multiput(7500,4800)(0,-400){12}{\line(0,300){100}}
\drawline\fermion[\E\REG](14000,5000)[10000]
\drawloop\gluon[\N 8](16500,7000)
\multiput(19000,4800)(0,-400){12}{\line(0,300){100}}
\drawline\fermion[\E\REG](28000,5000)[10000]
\drawline\photon[\NE\REG](29900,4800)[5]
\multiput(33000,4800)(0,-400){12}{\line(0,300){100}}
\drawline\gluon[\NW\FLIPPED](35900,5000)[2]
\end{picture}
\end{center}
\begin{center}
\begin{picture}(40000,8000)
\small
\drawline\fermion[\E\REG](0,5000)[10000]
\drawline\photon[\NW\FLIPPED](8200,4800)[5]
\multiput(5000,4800)(0,-400){12}{\line(0,300){100}}
\drawline\gluon[\NE\REG](2100,5000)[2]
\drawline\fermion[\E\REG](14000,5000)[10000]
\drawline\gluon[\NE\REG](16500,4900)[1]
\multiput(16500,4800)(0,-400){12}{\line(0,300){100}}
\drawline\photon[\NW\FLIPPED](20300,5000)[3]
\drawline\fermion[\E\REG](28000,5000)[10000]
\multiput(35500,4800)(0,-400){12}{\line(0,300){100}}
\drawline\gluon[\NW\FLIPPED](35500,4900)[1]
\drawline\photon[\NE\REG](31700,5000)[3]
\end{picture}
\end{center}
\begin{description}
\small
\baselineskip=12pt
\item[Figure 2:] Feynman diagrams which contribute to the gravitational
renormalization of the fermion-fermion-photon vertex at one-loop order.
Straight lines denote the fermion fields, wavy lines depict the spin-connection
fields $\omega_\mu^a$, spiral lines represent the dreibein fields $e_\mu^a$,
and dashed lines denote the background photon field $A_\mu$.
\end{description}
\end{figure}

Thus, as anticipated on general grounds from the discussion of the previous two
Sections, the low-energy limit of the parity-odd part of the vertex function
(and hence the renormalization of the spin) at one-loop order is determined
solely by the interaction of the fermion fields with the spin-connection
$\omega_\mu^a$. The triangle diagram for the exchange of one spin-connection
field in Fig. 2 contributes the term
\beq
\Gamma_\mu^{[\omega]}(p,q)=-\frac{4\pi i}{k'}\int\frac{d^3k}{(2\pi)^3}~{\cal
W}_j^\nu S^{(0)}(p-q-k)\gamma_\mu S^{(0)}(p-k){\cal
W}_i^\lambda~\Omega_{\lambda\nu}^{ij}(k)
\label{triangom}\eeq
to the total one-loop irreducible vertex function, where
\beq
S^{(0)}(p)\equiv\left\langle\!\!\left\langle\bar\psi(p)\psi(-p)
\right\rangle\!\!
\right\rangle=i(p_\mu\gamma^\mu-m)^{-1}=\frac
i{p^2-m^2}\left(p_\mu\gamma^\mu+m\right)
\label{fermprop}\eeq
is the free fermion propagator in momentum space. We can simplify the terms in
(\ref{triangom}) by successively applying the (2 + 1)-dimensional gamma-matrix
identity (\ref{2+1gammaid}) and the on-shell conditions
$p_\mu\gamma^\mu=(p-q)_\mu\gamma^\mu=-m$, $p^2=(p-q)^2=m^2$ and $p\cdot
q=q^2/2$ for the external particles (imagining, as mentioned above, that the
vertex function (\ref{triangom}) appears contracted between the two Dirac
spinors $\bar u(p-q)$ and $u(p)$). This reduces (\ref{triangom}) to the sum of
a piece proportional to $\gamma_\mu$ and a piece proportional to
$\sigma_{\mu\nu}$, as described in (\ref{formfacdef}). We are interested in the
parity-odd contributions, i.e. the terms in (\ref{triangom}) proportional to
$\epsilon_{\mu\nu\lambda}$. To extract these pieces, we simplify the vertex
function using the spin-connection propagator identities listed in the
Appendix. After a long and tedious algebraic calculation, we arrive at
\beq\new{\begin{array}{lll}
\Gamma_o^{[\omega]}(q^2)&=&\frac{8\pi
im}{k'M^2q^2(q^2-4m^2)}\int\frac{d^3k}{(2\pi)^3}~\frac1{(k^2-2k\cdot
p)(k^2-2k\cdot(p-q))(k^2-M^2)}\\&
&~~~~~\times\left\{\left(\epsilon_{\mu\nu\lambda}k^\mu p^\nu
q^\lambda\right)^2\left[\left(\frac3{2m}+\frac1M\right)\left(k^2-M^2
\right)-M\right]\right.\\&
&~~~~~+\frac{q^2}4\left(4m^2-q^2\right)\left[M(k^2-M^2)
-\frac1{2M}\left(3M^4+(k^2)^2\right)+\frac{M^2}{4m}[k\cdot(q-2p-k)]
\right.\\& &~~~~~\left.+\frac1{4m}k^2[k
\cdot(10p-3q-3k)]\right]+\left(m^2(k\cdot q)-\frac{q^2}2(k\cdot
p)\right)\left[m(3k^2+M^2)\right.\\& &~~~~~+\frac M
2[k\cdot(5k-p-7q)]-\frac{k^2}{2M}
[k\cdot(5p+3q)]-\frac3{2M}\left(M^4-2(k^2)^2\right)\\&
&~~~~~\left.+\frac{k^2}m[k\cdot(5p-2q
-3k)]+\frac{M^2}m[k\cdot(q-2p-k)]\right]\\& &~~~~~
+\frac{q^2}2[k\cdot(q-2p)]\left[\frac{3M}4\left(3M^2-2k^2\right)
-m(k^2+6M^2)+(k\cdot p)\left(\frac{k^2}M-\frac{7M}2\right)\right.\\&
&\left.\left.~~~~~-\frac{9(k^2)^2}{4M}+\frac{k^2}{2m}[k\cdot(q-3p+3k)]
+\frac{M^2}{2m}[k\cdot(3q-5p+k)]\right]\right\}\end{array}}
\label{gamomoint}\eeq

The (2 + 1)-dimensional momentum space integrations in (\ref{gamomoint}) are
absolutely convergent, and after some manipulation it is possible to cancel the
higher powers of the loop momentum in the Feynman integral (\ref{gamomoint})
with the denominator factors in its integrand. For example, the trivial
identity
\beq
2k\cdot p=-(k^2-2k\cdot p)+(k^2-M^2)+M^2
\label{trivid}\eeq
reduces the tensorial rank of a given Feynman integration in (\ref{gamomoint})
by 1 leaving a series of lower-rank integrals, most of which have fewer
denominator factors in their integrands. Then all the loop momentum
integrations can be carried out using the Feynman parametrizations
\beq
\frac1{ab}=\int_0^1dx~\frac1{[(1-x)b+xa]^2}~~~~~,~~~~~\frac1{abc}=2\int_0^1dx~
\int_0^xdy~\frac1{[ay+b(x-y)+c(1-x)]^3}
\label{feynpars}\eeq
and the (2 + 1)-dimensional Feynman integral identities
\beq\new{\begin{array}{l}
I^{(r)}(\alpha-p^2)\equiv\int\frac{d^3k}{(2\pi)^3}~\frac1{(k^2+2k\cdot
p+\alpha)^r}=\frac{\Gamma(r-3/2)}{8\pi^{3/2}\Gamma(r)}
\frac1{(\alpha-p^2)^{r-3/2}}\\\int\frac{d^3k}{(2\pi)^3}~\frac{k^\mu}
{(k^2+2k\cdot p+\alpha)^s}=-p^\mu
I^{(s)}(\alpha-p^2)\\\int\frac{d^3k}{(2\pi)^3}~\frac{k^\mu k^\nu}{(k^2+2k\cdot
p+\alpha)^s}=\left(p^\mu
p^\nu+\frac{\alpha-p^2}{2s-5}~\eta^{\mu\nu}\right)I^{(s)}(\alpha-p^2)
\end{array}}
\label{2+1feynints}\eeq
which hold for $2r>3$ and $2s>5$. Using these identities and some further
algebraic manipulations, after a long computation we arrive at the exact result
for the spin-connection form factor
\beq\new{\begin{array}{lll}
\Gamma_o^{[\omega]}(q^2)&=&\frac
m{336k'M^2(q^2-4m^2)}\left\{32M\left(4m^2-q^2\right
)\left[\frac{iF^{(1)}(q^2)}{\sqrt{-q^2}}-\frac{M^2}2\left(\frac M
m-1\right)F^{(2)}(q^2)\right]\right.\\& &-8M\left(12m^2q^2-32m^4+96Mm^3-24M
mq^2-\frac{M(q^2)^2}m+19M^3m-\frac{6M^3q^2}m\right.\\&
&\left.-22M^2m^2+7M^2q^2\right)F^{(3)}(q^2)-\frac{8M}{q^2}
\left(4(q^2)^2-16m^2q^2+64Mm^3-16mMq^2\right.\\&
&\left.-14M^2q^2+46m^2M^2-\frac{48m^2(q^2)^2}{M^2}-32M^3m
+\frac{15M^3q^2}m\right)F^{(4)}(q^2)\\& &+8|m|\left(\frac{2160m^2}
M-108m+\frac{185q^2}m-\frac{124q^2}M\right)+\frac{2M}m\left(4+\frac{23M}
m\right)\left(9q^2-9Mm\right.\\& &\left.-4m^2\right)-\frac
i{\sqrt{-q^2}}\left(\frac{34(q^2)^2}m-92mq^2-\frac{3252m^2q^2}M-
\frac{35(q^2)^2}M-\frac{288m^4}M-400m^3\right.\\& &\left.-8mq^2-48M
m^2+100mM^2+18Mq^2-\frac{25M^2q^2}m\right)\log\left(\frac{2|m|-i
\sqrt{-q^2}}{2|m|+i\sqrt{-q^2}}\right)\\& &\left.
+\frac{4M}m\left(\frac{9M^2q^2}{m^2}-\frac{9M^3}m-4M^2+32m^2-8q^2
\right)\log\left(1+\frac{2m}M\right)\right\}\end{array}}
\label{spintriangform}\eeq
where $F^{(j)}(q^2)$ are the Feynman parametric integrals
\beq\new{\begin{array}{l}
F^{(1)}(q^2)=\int_0^1dx~\log\left(\frac{3\sqrt{-q^2}x+2i\sqrt{M^2(1-x)+
(m^2+2q^2)x^2}}{\sqrt{-q^2}x+2i\sqrt{M^2(1-x)+m^2x^2}}\right)\\F^{(2)}(q^2)=
\int_0^1dx~\frac x{4M^2(1-x)+(4m^2-q^2)x^2}
\\~~~~~~~~~~~~~~~~~~~~~~~\times\left(\frac3{\sqrt{M^2(1-x)+(m^2+2q^2)x^2}}
-\frac1{\sqrt{M^2(1-x)+m^2x^2}}\right)\\F^{(3)}(q^2)=\int_0^1dx~
\frac{x^2}{4M^2(1-x)+(4m^2-q^2)x^2}\\~~~~~~~~~~~~~~~~~~~~~~~\times
\left(\frac3{\sqrt{M^2(1-x)+(m^2+2q^2)x^2}}-\frac1{\sqrt{M^2(1-x)+m^2x^2}}
\right)\\F^{(4)}(q^2)=\int_0^1dx~\frac1{4M^2(1-x)+(4m^2-q^2)x^2}
\\~~~~~~~~~~~~~~~~~~~~~~~\times\left(\frac{M^2(1-x)+(q^2+2m^2)x^2}
{\sqrt{M^2(1-x)+m^2x^2}}-2\sqrt{M^2(1-x)+(m^2+2q^2)x^2}
\right)\end{array}}
\label{feynparints}\eeq
The integrated forms of $F^{(j)}(q^2)$ are quite complicated and not very
informative for generic values of $q^2$. However, in the momentum regimes of
interest to be discussed below, they can be evaluated in terms of simple
algebraic forms.

The vertex function (\ref{spintriangform}) is finite in the infrared limit and
at $q^2\to0$ we find
\beq\new{\begin{array}{lll}
\Gamma^{[\omega]}_o(0;M)&=&-\frac1{672k'M^2m}\left\{M\log
\left(1+\frac{2m}M\right)\left[\frac{65M^2}m-448M-192m-\frac{62M^3}{m^2}
\right]\right.\\& &+\frac M{M+2m}\left[256m^2-896M
m-\frac{88M^3}m+84M^2\right]\\& &\left.+70M
m+\frac{8784m^3}M-232m^2-\frac{212M^3}m-178M^2\right\}\end{array}}
\label{spinform0}\eeq
where for definiteness we have taken $m>0$. Note that the finiteness of the
irreducible vertex function at $q^2=0$ confirms the general expectations of the
infrared finiteness of topologically massive gravity in the Landau gauge
\cite{desyang,kesz}. There are two different mass regimes of interest. The
first is the regime of a heavy graviton field with mass $M>m$ and the other is
that of a heavy fermion field of mass $m>M$. The topological Einstein limit of
the gravity theory is the extreme first regime when $M\to\infty$, and taking
this limit in (\ref{spinform0}) yields
\beq
\lim_{M\to\infty}\Gamma_o^{[\omega]}(0;M)=-\frac1{2k'm}
\label{spinmagmom}\eeq
Comparing with (\ref{magmomdef}) and incorporating the tree-level (bare) vertex
function (see (\ref{gordondecomp})) we find that the total magnetic moment
including the one-loop anomalous contribution is
\beq
\mu^{(1)}=\frac2{2m}\left(\frac12+\frac1{4k'}\right)
\label{magmom1loop}\eeq
which identifies the total one-loop order gravitationally dressed spin
$\Delta_{1/2}^{(1)}(k')=1/2+1/4k'$, as predicted by the KPZ formula. Thus in
the limit $M\to\infty$ where topologically massive gravity becomes topological
Einstein gravity, the magnetic moment structure generated via the gravitational
renormalizations by the spin-connection field yield another definition of the
induced spin of the charged spinor fields which coincides with the predictions
of quantum Liouville theory.

In the extreme second regime where $M\ll m$ we find
\beq
\lim_{M\to0}\Gamma_o^{[\omega]}(0;M)=-\frac{183}{14k'm}\left(\frac mM\right)^3
\label{spingam0}\eeq
Just like the tree-level interaction amplitudes, the one-loop vertex function
diverges in the limit $M\to0$ when the topological Einstein term in the gravity
action becomes irrelevant and the gravity theory is equivalent to an $SO(2,1)$
Chern-Simons gauge theory (and possesses local conformal symmetry). In general,
the $M$-dependent vertex function (\ref{spinform0}) determines an anomalous
dimension $\Delta(M/m)$ which depends on the mass scale of the fermion-coupled
topologically massive gravity theory. Since the natural bulk scale
$M^2\propto\kappa^2$ of three-dimensional gravity coincides with the natural
scale $\Lambda$ of the induced Liouville theory, this can be taken as a
definition of scale dependent KPZ conformal dimensions, which are otherwise not
predicted by the two-dimensional KPZ theory.

After some calculation we also find that $\Gamma_o^{[\omega]}(q^2)\to0$ at
$q^2\to\infty$. This means that the bare magnetic moment (and hence the bare
spin $\frac12$) of the fermion fields is not renormalized by the gravitational
interaction at short distance scales. On the other hand, the long-range effects
of the gravitational field do lead to an anomalous magnetic moment. However, it
is only in the topological limit $M\to\infty$ that we obtain the expected
Liouville dressed spin and the correct gyromagnetic ratio $g=2$ for the fermion
fields. For finite values of $M$, the vertex function (\ref{spinform0}) can
also be thought of as defining a scale dependent gyromagnetic ratio
$g(\Lambda)=g(M/m)$ for the fermion fields which reduces to the canonical value
$g=2$ at large-distance (i.e. topological) scales. The observables of
matter-coupled topologically massive gravity in this infrared limit are
associated with those of string theory in the critical dimension when the
induced gravitational anomaly vanishes. However, since the anomalous spin in
the fermion-coupled topologically massive gravity theory was shown in the
previous Section to also coincide with the KPZ formula in the ultraviolet limit
$M\to0$, the qualitative features of our results are in effect independent of
the topological graviton mass scale. They do not depend on which regime of the
theory we are in and therefore hold for the full induced Liouville gravity
theory as well. This is the important and somewhat surprising feature of the
correspondence here between the two- and three-dimensional theories. It is
interesting that these two regimes of topologically massive gravity actually
represent two different symmetry points in the moduli space of
three-dimensional quantum gravity theories. Either symmetry group leads to the
appropriate renormalization of spin predicted by the KPZ theory. As we discuss
in the next Section, the more natural appearence of the KPZ weights at
Poincar\'e-invariant point corresponds to the fact that at this point in the
moduli space the three-dimensional quantum gravity theory coincides in some
sense with topological gravity.

It would be interesting to extend the above cumbersome calculation (as well as
those of the previous Section) to higher-loop orders. This would serve as a
non-trivial check of the relationship between quantum Liouville theory and
topologically massive gravity. It would also illustrate how the different
gravity fields of the three-dimensional theory conspire to correctly
renormalize spin beyond that which can be predicted based on a naive tree-level
calculation. Such calculations would also provide further information about the
seemingly mysterious scale dependences of the basic observables, for instance
if there is any renormalization giving $g-2\neq0$ at higher-loop orders and
hence leading to other gravitational renormalizations of the (anomalous)
fermion magnetic moment. It would be interesting to use this scale dependence
to further illustrate the precise roles of the symmetry groups of the
three-dimensional gravity theory in relation to the induced two-dimensional
quantum gravity theory.

\section{Induced Spin and Phases of Three-dimensional Quantum Gravity}

In this final Section we shall discuss some puzzles concerning induced spin in
topologically massive gravity, and present some conjectures about how these
unusual features of the three-dimensional theory could be related to different
phases of the Liouville theory (\ref{liouville}). Our first observation
concerns the other branch of the KPZ scaling relations (\ref{KPZtransf}), which
is related to the solution (\ref{KPZexp}) that vanishes at $\Delta_0=0$ by
\beq
\Delta_+=-(k'+1)-\Delta_-
\label{Delta+}\eeq
This solution of the KPZ formula diverges in the asymptotic limit
$k'\to\infty$. As is well-known, the KPZ theory is only valid for
two-dimensional quantum gravity coupled to matter fields of central charge
$c\leq1$, in which case the canonical choice of dressed scaling dimension
$\Delta_-$ is taken. The other choice (\ref{Delta+}) for the dressed weights
corresponds to changing the gravitational dressing in the tachyon operator term
in (\ref{liouville}) from $\e^{\alpha_+\phi}$ to $\e^{\alpha_-\phi}$, where
\beq
4\alpha Q=-2-\alpha^2
\label{alphaQdef}\eeq
It has been argued in \cite{kleb} that this change of branch in the KPZ scaling
relations effectively couples two-dimensional quantum gravity to matter fields
of central charge $c>1$, which are to be thought of as embedding the
world-sheet into a ``crumpled" phase, rather than a smooth phase. In this phase
the surfaces which dominate the string partition function are fragmented into
trees of ``baby universes", in each of which the usual $c<1$ behaviour is
exhibited. This second phase of two-dimensional quantum gravity is known as the
branched polymer phase, wherein the typical objects that contribute to the
string partition function at large distance scales are tree-like objects,
rather than the conventional two-dimensional surfaces which dominate in the
stringy phase.

{}From a dynamical point view it is interesting that the perturbative regime of
topologically massive gravity automatically selects the branch $\Delta_-$ of
the KPZ equations. In fact, as discussed in \cite{a-cksgrav}, the perturbative
approach that we have presented can only yield this branch, because the
effective coupling constant of the three-dimensional theory is $1/k'$. This
same sort of argument is used in quantum Liouville theory to select this
branch, because it is only for that choice that the semi-classical limit exists
in the weak-field approximation ($c\gg1$). In particular, it is impossible to
reach the conformally-invariant point in the parameter space of
three-dimensional quantum gravity ($k'\to\infty$) with the branch
(\ref{Delta+}). However, there is another sort of induced spin in topologically
massive gravity which could be related to this other phase of the
two-dimensional gravity theory. This induced spin is due to a purely
topological feature of the effective three-dimensional spacetime and it is
directly related to the gravitational analog of the Aharonov-Bohm effect
\cite{desmc} due to the mass-energy generated by sources. This effect is most
natural in three-dimensional gravity, where the vacuum Einstein equations
$R_{\mu\nu}=0$ imply that locally the curvature vanishes, i.e.
$R_{\mu\nu\lambda\rho}=0$. Thus outside of the support of matter fields the
spacetime is flat, but, as shown in \cite{3dconical}, globally the theory can
exhibit non-trivial effects. In contrast to the four-dimensional case, the
field equations imply exactly the classical equations of motion of particle
singularities. As discussed in \cite{mazur}, a source of mass $m$ and spin
$\Delta_0$ gives a helical time structure for the effective spacetime which can
lead to interesting quantum mechanical effects. Furthermore, the spacetime has
a locally flat conical geometry which shifts the angular momentum spectrum
leading to an induced spin. Such a spacetime is characterized by a time
interval $t\in[0,4\kappa^{-1}\Delta_0]$ and a plane polar angle range
$\phi\in[0,2\pi-4\kappa^{-1}m]$. Thus a point spinning particle induces a
non-static spacetime with a curvature singularity at the origin (corresponding
to a non-trivial holonomy about the origin) and also a torsion singularity. It
is possible to measure the mass $m$ (per unit length) through the gravitational
Aharonov-Bohm effect. This effect is of order $\kappa^{-1}m$.

In topologically massive gravity, the Einstein field equations are replaced by
the Einstein-Cotton equations
\beq
\sqrt{g}~R_{\mu\nu}+M^{-1}C_{\mu\nu}=-\kappa^{-1}\sqrt{g}~T_{\mu\nu}
\label{eincoteqs}\eeq
where $T_{\mu\nu}$ is the energy-momentum tensor of a matter source and
\beq
C_{\mu\nu}=\epsilon_{\mu\lambda\rho}\nabla^\lambda\left(R^\rho_\nu-\frac14
\delta^\rho_\nu R\right)
\label{cotton}\eeq
is the trace-less, symmetric Cotton tensor density which is conserved
($\nabla^\mu C_{\mu\nu}=0$) and conformally-invariant ($C_{\mu\nu}$ is the
three-dimensional analog of the Weyl tensor). The static and circularly
symmetric solutions of the topologically massive gravity field equations
(\ref{eincoteqs}) for a point particle of mass $m$ and spin $\Delta_0$ are
\cite{desmc}
\beq
ds^2\equiv h_{\mu\nu}(x)dx^\mu\otimes dx^\nu=N(r)dt\otimes dt+\varphi(r)d\vec
x\otimes d\vec x+dt\otimes d\vec x\times\vec\partial W(r)
\label{spacetimepar}\eeq
where $h_{\mu\nu}\equiv g_{\mu\nu}-\eta_{\mu\nu}$ and
\beq\new{\begin{array}{l}
W(r)=-\frac{M^{-1}}{2\pi\kappa}\left(m+M\Delta_0\right)\left(\log
(Mr)+K_0(Mr)\right)\\\varphi(r)=\frac1{2\pi\kappa}\left(m+M\Delta_0\right)
K_0(Mr)+\frac m{\pi\kappa}\log(M
r)\\N(r)=\frac1{2\pi\kappa}\left(m+M\Delta_0\right)K_0(Mr)\end{array}}
\label{tmgexplsols}\eeq
with $K_0(x)$ the irregular modified Bessel function of order 0 which has the
asymptotic behaviours $K_0(x)\sim-\log x$ for $x\to0$ and $K_0(x)\sim
x^{-1/2}\e^{-x}$ for $x\to\infty$. These solutions are only valid for a
weak-field metric (i.e. in the linearized approximation to the full non-linear
theory).

The spatial part of the metric at infinity (corresponding to the infrared limit
of topologically massive gravity) is
\beq
\lim_{r\to\infty}\varphi(r)=\frac m{\pi\kappa}\log(Mr)
\label{spaceasympt}\eeq
so that the solution (\ref{tmgexplsols}) represents an asymptotically conical
space whose angular deficit range exceeds $2\pi$. An exact solution in the full
non-linear theory with this asymptotically conical structure is possible only
when $m+M\Delta_0=0$, in which case the above solution has the same form as
that generated by a spinless point mass in pure Einstein gravity. In the
linearized approximation, it is possible to construct the Noether generators
using the above field equations, and one finds in the adiabatic limit that the
canonical angular momentum of the particle is shifted by an amount
\beq
\Delta_{\rm ind}=-\frac{m^2}{32\pi^2\kappa^2}k'=-\frac2{k'}\left(\frac
mM\right)^2
\label{indspin}\eeq
which corresponds to an induced spin of the particle generated by its
interaction with the topologically massive gravity field. The quantity $m/M$
measures the amount of gravitational flux generated by the point mass $m$. Thus
the induced spin in this non-perturbative, weak-field approach is scale
dependent and has the same qualitative properties as the spin determined by the
vertex function (\ref{spingam0}) in the ultraviolet regime of the topologically
massive gravity theory. It is only on mass scales of the order of the
topological graviton mass that we obtain an induced spin resembling the
Liouville dressed conformal dimension arising from the perturbative regime of
the three-dimensional theory.

The induced spin (\ref{indspin}), which is non-vanishing even for spinless
particle sources ($\Delta_0=0$) will be more complicated in the full non-linear
theory, but it is always determined as some non-perturbative dynamical effect
in the theory. This induced spin comes from a gravitational analog of the
Aharonov-Bohm effect since it is determined by the gravitational flux generated
by the mass $m$. The linearized approximation in this approach is the analog of
the tree-level perturbative calculations that we presented earlier. The fact
that the induced spin (\ref{indspin}) is non-zero for spinless particles
suggests that it could be related to the non-perturbative branch $\Delta_+$ of
the KPZ scaling relations discussed above. It would be interesting to explore
this potential relationship further, and also to examine how to relate to
topologically-induced spins (\ref{indspin}), which are induced by identical
particle exchange in the asymptotic, field-free region of the spacetime, with
the perturbative spins that we derived before, which arise from identical
particle exchange where the particles interact gravitationally with each other
in a flat space-time (without conical singularities). The charged particles act
as sources for the gravitons which renormalize the bare spin of the particles.
It is unclear how to relate the low-energy quantum field theoretical amplitudes
obtained in this Paper to the classical conical structure of spacetime
\cite{a-cksgrav}. Various solutions to the field equations exhibiting such
singularity structures have been obtained recently in \cite{clement}, and it
would be interesting to connect these structures with the polymer structures
dominating $c>1$ string theory.

The topological nature of the induced spin (\ref{indspin}) also suggests that
it could be related to some more direct topological property of
three-dimensional gravity. One candidate is three-dimensional topological
gravity, where the dreibein and spin-connection fields are taken as independent
variables ($S_\lambda\equiv0$ in (\ref{TMGaction})). The classical equations of
motion of this theory are $R=\nabla e^a=0$, so that at tree-level the induced
spins in topological gravity will coincide with those above in topologically
massive gravity.  This theory can be written as a topological $ISO(2,1)$
Chern-Simons gauge theory and it was shown in \cite{cho} that a similar
solution to that given above is obtained {\it exactly} in this full non-linear
theory (and not just when $m+M\Delta_0=0$). This theory has no dynamical
degrees of freedom, and thus it may be related to the $c>1$ phase of the
quantum Liouville theory. Notice that in three-dimensional topological gravity
one expects to obtain scattering amplitudes with kinematical poles
characteristic of the Aharonov-Bohm effect, which is why the KPZ weights were
more naturally identified before using the Poincar\'e symmetry of the
three-dimensional quantum field theory.

The other possibility is that the $\Delta_+$ branch of the KPZ formula is
reached in the topological phase of topologically massive gravity
\cite{kogan1}, i.e. the phase wherein $\langle e_\mu^a\rangle=0$ and there is
no background spacetime. In this phase one cannot diagonalize the quadratic
form of the bulk part of the action to find the propagators
\cite{kogan1,a-cks}, because there are only two gauge groups in the model
(diffeomorphisms and local $SO(2,1)$ rotations) while there are three fields
($\beta$, $\omega$ and $e$) that require gauge-fixing. Thus the quadratic
approximation does not exist and perturbation theory breaks down. Furthermore,
when $\langle e_\mu^a\rangle=0$ the local $SO(2,1)$ symmetry is unbroken and
both space-time and tangent space rotations individually preserve the symmetry
of the ground state. It could be that this phase of the gravity theory
(\ref{TMGaction}) is intimately related to $c>1$ string theory. Note that there
are actually several such topological phases, depending on how degenerate the
dreibein field is. The intermediate phases occur when $\langle
e_\mu^a\rangle\neq0$, but $\det_{\mu,a}[\langle e_\mu^a\rangle]=0$. The number
of different ways that this degeneration can occur depends on the topology of
the underlying 3-manifold $\cal M$. It would be interesting to use the
topological membrane approach in this way to model the fractal baby universe
structure of the branched polymer phase of two-dimensional quantum gravity.

Some evidence for such topological interpretations lies in the fact that for
the $\Delta_-$ gravitational dressing, the puncture operator on the worldsheet
$\Sigma$ is the local tachyon vertex operator $\e^{\alpha_+\phi}$ which creates
microscopic loops on the worldsheet. This corresponds to the perturbative phase
of the three-dimensional theory in which there are local, propagating graviton
degrees of freedom which are exchanged by charged particles moving in the bulk
$\cal M$. When the gravitational dressing is changed to the $\Delta_+$ branch,
it is the non-local, smeared tachyon vertex operator $\int_\Sigma
d^2z~\e^{\alpha_-\phi}$ which creates macroscopic loops on $\Sigma$. This is a
global area form which corresponds to a topological phase of the model in which
there are no propagating degrees of freedom at all. The distinction in
two-dimensions between microscopic states corresponding to local operators and
macroscopic states follows from the fact that the metric is a dynamical
variable of the quantum Liouville theory. The dressing of the tachyon operator
by a macroscopic state is not merely a local disturbance to the surface,
because it creates a macroscopic hole and tears the surface apart. Integrating
over all such contributions in the string partition function leaves all but a
microscopic (of the order of the cutoff) fraction of the worldsheet with holes.
Thus the surface deteriorates and the resulting model is in the branched
polymer phase discussed above. From the three-dimensional point of view, the
integrated tachyon vertex operators correspond to smeared-out particle
worldlines (see Section 1) wherein the graviton exchanges cannot be described
perturbatively. Thus the three-dimensional description of quantum Liouville
theory has the potential of even providing a geometrical approach to the
mysterious $c>1$ phase of string theory. It would be interesting to develop
some alternative non-perturbative approach, such as a connection with the
conical structure of the three-dimensional spacetime (and hence the
gravitational Aharonov-Bohm effect), to describe this phase.

\section*{Acknowledgements}

We thank J. Wheater for helpful discussions. The work of R.J.S. was supported
in part by the Natural Sciences and Engineering Research Council of Canada.

\setcounter{section}{0}
\setcounter{subsection}{0}
\addtocounter{section}{1}
\setcounter{equation}{0}
\setcounter{equnum}{0}
\renewcommand{\thesection}{\Alph{section}}

\section*{Appendix \ \ \ Gravitational Propagator Identities}

In this Appendix we summarize some identities for the various gravitational
propagators that are used throughout this Paper. The relevant formulas for the
spin-connection propagator are as follows:
\beq\new{\begin{array}{l}
\Omega^{ij}_{i\nu}(q)=-\frac{1}{2M}\Lambda^j_\nu(q)+i\epsilon^j_{~\nu
\lambda}q^\lambda\frac{q^2+M^2}{2M^2(q^2-M^2)}\\\Omega^{ij}_{\mu
j}(q)=-\frac{1}{2M}\Lambda^i_\mu(q)-i\epsilon^i_{~\mu
\lambda}q^\lambda\frac{q^2+M^2}{2M^2(q^2-M^2)}\\\Omega^{ij}_{ij}(q)=
\frac{1}{2M}\left(\frac{q^2}{M^2}-3\right)\end{array}}
\label{spinpropids}\eeq

\beq\Omega_{\lambda
i}^{ij}(q)=\frac1{q^2-M^2}\left(\frac{q^2-3M^2}{2M}\Lambda_\lambda^j(q)
+i\epsilon_{~\lambda\rho}^{j}q^\rho\right)+\frac1{2M^3}\left(2M^2
\delta_\lambda^j-\frac{2(q^2-3M^2)}{M^2}q_\lambda q^j\right)
\label{spinpropidk1}\eeq

\beq
\Omega_{j\nu}^{ij}(q)=\frac1{q^2-M^2}\left(\frac{q^2-3M^2}{2M}
\Lambda_\nu^i(q)-i\epsilon_{~\nu\rho}^{i}q^\rho\right)+\frac1{2M^3}
\left(2M^2\delta_\nu^i-\frac{2(q^2-3M^2)}{M^2}q^iq_\nu\right)
\label{spinpropidk2}\eeq

\beq
\Omega_{ji}^{ij}(q)=-\frac1{M^3(q^2-M^2)}\left(6M^2(q^2-2M^2)+
\frac{(q^2)^2}{M^2}(q^2-3M^2)\right)
\label{spinpropidk3}\eeq

\beq
[\eta_{ij}\Omega^{ij}_{\lambda\nu}(q)]^{\rm
odd}=\frac{i(q^2-2M^2)}{M^2(q^2-M^2)}\epsilon_{\lambda\nu\rho}q^\rho
{}~~~~~~,~~~~~~[\eta^{\mu\nu}\Omega_{\mu\nu}^{ij}(q)]^{\rm odd}=\frac
i{q^2-M^2}\epsilon^{ij\lambda}q_\lambda
\label{spinpropidk4}\eeq

\beq
q^\lambda\Omega_{\mu\lambda}^{ij}(q)=-\frac1{2M^3}\left(q_\mu
q^iq^j+M^2\delta_\mu^iq^j+2(q^2-M^2)\eta^{ij}q_\mu+2q^2\delta^j_\mu
q^i+iM\epsilon^{i\lambda}_{~\mu}q_\lambda q^j\right)
\label{qspinpropid}\eeq

\beq
\epsilon^\mu_{~i\rho}\Omega_{\mu\nu}^{ij}(q)=\frac{q^\mu}{M(q^2-M^2)}\left
(\epsilon_{\mu~\rho}^{~j}q_\nu-\epsilon_{\mu\nu\rho}q^j\right)-\frac
i{M^2}\eta_{\rho\nu}q^j
\label{spinpropepid}\eeq

\beq
\epsilon_{~j\rho}^\nu\Omega_{\mu\nu}^{ij}(q)=\frac
i{M^2}q^i\eta_{\mu\rho}-\frac1M\epsilon_{\mu~\rho}^{~i}+\frac1{2M^3}
\left(\epsilon^{\nu i}_{~~\rho}q_\mu
q_\nu-3\epsilon_{\mu~\rho}^{~\nu}q^iq_\nu\right)
\label{spinpropidk5}\eeq

\beq
[\epsilon_{\mu i}^{~~\nu}\Omega_{\lambda\nu}^{ij}(q)]^{\rm
odd}=\frac1{2M(q^2-M^2)}\left((3M^2-q^2)\epsilon_{\mu\lambda}^{~~j}
+\frac{2(q^2-2M^2)}{M^2}\left[\epsilon_{\mu
i}^{~~j}q^iq_\lambda+\epsilon_{\mu\lambda}^{~~\nu}q_\nu q^j\right]\right)
\label{spinpropidk5a}\eeq

\beq
[\epsilon_{\mu ij}\Omega_{\lambda\nu}^{ij}(q)]^{\rm
odd}=\frac{3M^2-q^2}{2M(q^2-M^2)}\epsilon_{\mu\lambda\nu}-
\frac1{M(q^2-M^2)}\left(\epsilon_{\mu i\nu}q^iq_\lambda+\epsilon_{\mu\lambda
j}q^jq_\mu\right)
\label{spinpropidk7}\eeq

\beq
[\epsilon_\mu^{~\lambda\nu}\Omega_{\lambda\nu}^{ij}(q)]^{\rm
odd}=\frac{3M^2-q^2}{2M(q^2-M^2)}\epsilon_\mu^{~ij}-\frac1{M(q^2
-M^2)}\left(\epsilon_\mu^{~\lambda j}q_\lambda q^i+\epsilon_\mu^{~i\nu}q_\nu
q^j\right)
\label{spinpropidk8}\eeq

\beq
[\epsilon_{\mu~j}^{~\lambda}\Omega_{\lambda\nu}^{ij}(q)]^{\rm
odd}=\frac{q^2+M^2}{2M(q^2-M^2)}\epsilon_{\mu~\nu}^{~i}-
\frac1{M(q^2-M^2)}\left(\epsilon_{\mu~\nu}^{~\lambda}q_\lambda
q^i+\epsilon_{\mu~j}^{~i}q_\nu q^j\right)
\label{spinpropidk9}\eeq

\beq
[\epsilon^\lambda_{~i\beta}\epsilon^\nu_{~j\mu}
\Omega_{\lambda\nu}^{ij}(q)]^{\rm odd}=-\frac i{M^2}
\epsilon_{\beta j\mu}q^j
\label{spinpropidk6}\eeq
where $[~\cdot~]^{\rm odd}$ denotes the parity-odd part of the given tensor
function, i.e. the piece proportional to $\epsilon_{\mu\nu\lambda}$. We have
also made use of the following identities for the graviton propagator:
\beq
D^{ij}_{i\nu}(q)=D^{ji}_{\nu
i}(q)=\frac{i}{\kappa}\left(\frac{q^2+M^2}{2q^2(q^2-M^2)}\delta^{\perp
j}_\nu(q)+iM\frac{\epsilon_{\nu~\lambda}^{~j}q^\lambda}{q^2(q^2-M^2)}
\right)~~~,~~~D^{ij}_{ij}(q)=\frac{i}{\kappa}\frac{q^2+M^2}{q^2(q^2-M^2)}
\label{gravpropids}\eeq

\end{document}